\documentstyle[12pt]{article}
\begin{document}
\setlength{\baselineskip}{0.7cm}

\title{Initial-conditions problem for a Chiral Gross-Neveu system}

\author{P.L. Natti \thanks{Present address : Departamento de
F{\'{\i}}sica, Universidade Estadual de Londrina, Paran\'a,
Brasil}\\ Instituto de F{\'{\i}}sica Te\'orica, Universidade Estadual
Paulista,\\ Rua Pamplona,145 - 01405-900 S\~ao Paulo, S.P., Brasil\\
\\ A.F.R. de Toledo Piza \\ Instituto de F\'{\i}sica, Universidade de
S\~ao Paulo,\\ C.P. 66318, 05389-970 S\~ao Paulo, S.P., Brasil}

\maketitle
\begin{abstract}
 
A time-dependent projection technique is used to treat the
initial-value problem for self-interacting fermionic fields. On the
basis of the general dynamics of the fields, we derive formal
equations of kinetic type for the set of one-body dynamical
variables. A nonperturbative mean-field expansion can be written for
these equations.  We treat this expansion in lowest order, which
corresponds to the Gaussian mean-field approximation, for a uniform
system described by the Chiral Gross-Neveu Hamiltonian. Standard
stationary features of the model, such as dynamical mass generation
due to chiral symmetry breaking and a phenomenon analogous to
dimensional transmutation, are reobtained in this context. The
mean-field time evolution of non-equilibrium initial states is
discussed.

\end{abstract}
\newpage

\section{Introduction}

Over the last two decades, interest in the initial-conditions problem
for field theoretical models has been kindled and sustained by the
needs of such apparently diverse areas as cosmology \cite{1} and
several branches of many-body physics, notably in connection with the
analysis of transient phenomena in the collision of complex nuclear
systems \cite{2}. In these contexts one typically tries to obtain and
solve equations describing the kinetic behavior of a particular,
``relevant'' subsystem or of a restricted set of ``relevant''
observables of a more comprehensive autonomous system. Such is the
case e.g. of the scalar driving field in the inflationary scenario of
the early universe and of one-body densities and certain correlation
functions in heavy-ion collisions. In general, the relevant properties
can be retrieved from appropriately constructed reduced density
operators in the Schr\"odinger picture, which will evolve nonunitarily
on the account of correlation effects involving different subsystems
\cite {3} \cite {4}.  The nonunitary effects will manifest themselves
through the dynamical evolution of the eigenvalues of the reduced
densities, so that particular subsystems will in general evolve in a
nonisoentropic manner \cite {3}.

The overwhelming complexity of such a picture is considerably reduced
whenever one is able to find physical grounds to motivate a
mean-field-like approximation which consists in assuming isoentropic
evolution of a relevant subsystem under effective, time-dependent
Hamiltonian operators for each subsystem \cite {4}. In this case the
dynamics of the subsystem density matrix can be formulated in terms of
a Liouville-von Neuman equation governed by an effective Hamiltonian
and studied e.g. from the point of view of the functional
field-theoretical Schr\"odinger picture, as proposed by Jackiw \cite
{4}.  Unfortunately the resulting problem still involves in general
nonlinear Hamiltonians, and cannot be solved without further
approximation.  In the field-theoretical context, this has been
implemented through the use of a Gaussian ansatz for the subsystem
density functional in the framework of a time-dependent variational
principle supplying the appropriate dynamical information, notably for
bosonic fields \cite{KeLi}. 

It is not difficult to see that this last approximation amounts to a
second mean-field approximation, now at the microscopic level of the
single-field, nonlinear, isoentropic effective dynamics. Actually, the
Gaussian ansatz, having the form of a exponential of a quadratic form
in the field operators, implies that many-point correlation functions
can be factored in terms of two-point functions.  This is well known
in the context of the derivation of the Hartree-Fock approximation to
the nonrelativistic many-body problem \cite {5}. This factorization
has been used by Chang \cite {6} to implement the Gaussian
approximation for the $\lambda \phi^{4}$ theory. The dynamics of the
reduced two-point density becomes then itself isoentropic, since
irreducible higher-order correlation effects are neglected.

The focus of this work is a reevaluation of this second mean-field
approximation, for fermionic fields, in terms of a time-dependent
projection approach developed earlier for the nonrelativistic nuclear
many-body dynamics by Nemes and de Toledo Piza \cite {7}. This approach
allows for the formulation of a mean-field expansion for the dynamics
of the two-point correlation function from which one recovers the
results of the Gaussian mean-field approximations in lowest order,
i.e., this approach permit to include and to evaluate higher dynamical
corrections effects to the simplest Gaussian mean-field
approximation. Moreover, the expansion is energy-conserving (for
closed system) to all orders \cite {8}. The resulting dynamical
equations acquire the structure of kinetic equations which eliminate
the isoentropic mean-field constraint describing the effective
dynamics of a selected set of observables \cite {7}. This approach was
recently applied for the solution of the self-interacting $\lambda
\phi^{4}$ theory in (1+1) dimensions \cite {9}. Lin and de Toledo Piza
find that the Gaussian mean-field approximation fails qualitatively and
quantitatively in the description of certain field variables.  These
failures are partially corrected by the collisional terms. Motivated
by success obtained in description of time evolution of a
off-equilibrium uniform boson (scalar field) system beyond Gaussian
mean-field approximation in quantum-field theoretical context, it
becomes interesting to study the fermion case in this approach. As a
first step towards this end we consider in this paper the
implementation of the Gaussian approximation to a self-interacting
system of fermions. This is done in a framework suitable for the
subsequent inclusion of collisional (correlation) corrections along
the lines developed in ref. \cite{9} for the self-interacting bosonic
field.

We consider, for simplicity and definiteness, the case of an
off-equilibrium, spatially uniform many-fermion system described by
Chiral Gross-Neveu model \cite {10}. This is an interesting
non-trivial, renormalizable model for which many results are available
in the literature so that it offers suitable testing ground for the
proposed approach. On the basis of the general dynamics of the fields,
we derive equations of kinetic type for the set of one-body variables
in lowest order, which correspond to the Gaussian mean-field 
approximation. The detailed consideration of correlation corrections 
is deferred to future work.

An outline of the paper is as follows.  In Sec.II we obtain the
dynamical equations which describe the time evolution of a general
uniform fermion system. These equations are the groundwork for the
implementation of the time-dependent projection technique. This
technique and the approximation scheme are described in Sec.III. In
Sec.IV we implement in the quantum-field theoretical context the
projection technique and obtain in Gaussian mean-field (isoentropic)
approximation the equations which describe the effective dynamics of a
off-equilibrium spatially uniform (1+1) dimensional self-interacting
fermion system described by Chiral Gross-Neveu model \cite{10}. In
Sec.V we use the static solution of these equations in order to
renormalize the theory, leading to the well-known effective potential
obtained by Gross and Neveu using the $1/N$ expansion.  In this same
section, we show also that other static results which have been
discussed in the literature such as dynamical mass generation due to
chiral symmetry breaking and a phenomenon analogous to dimensional
transmutation can be retrieved from this formulation in a mean-field
approximation.  Finally we obtain and discuss numerical solutions for
non-equilibrium initial conditions. Sec.VI is devoted to a final
discussion and conclusions. Some points of a more technical nature are
discussed in the Appendices.

\section{Kinetics of a self-interacting fermionic field}

In this section, we shall describe a formal treatment of the kinetics
of a self-interacting quantum field. Although the procedure is quite
general, we will adopt the specific context of a spatially uniform
fermion system. We will illustrate all the relevant points of the
approach and cut down inessential technical complications. Features of
more general contexts are discussed in Ref. \cite {11}.

The idea of our approach is to focus on the time evolution of a set of
simple observables.  We argue that a large number of relevant physical
observables are one-body operators.  Consequently, the time evolution
of observables which involve field bilinear forms such as
$\bar\psi(x)\psi(x)$ , $\psi(x)\psi(x)$ , ....  is desirable. These
are the observables which are kept under direct control when one works
variationally using a Gaussian functional ansatz, and will therefore
be refered to as Gaussian observables.  In order to keep as close as
possible to the formulation appropriate for the many-body problem, we
work in fact with expressions which are bilinear in the creation and
annihilation parts of the fields in momentum space with periodic
boundary conditions in a spatial box of lenght $L$, defined in terms
of an expansion mass parameter $m$.  We begin by expanding the Dirac
field operators $\psi(x)$ and $\bar\psi(x)$ in Heisenberg picture as

\begin{eqnarray}
\psi(x)&=&\sum_{\bf k}\left(\frac {m}{k_{0}}\right)^{1/2}
\left[b_{{\bf k},1}(t)u_{1}({\bf k})\frac {e^{i{\bf kx}}}{\sqrt
{L}}+b^{\dag}_{{\bf k},2}(t)u_{2}({\bf k})\frac {e^{-i{\bf kx}}}{\sqrt
{L}}\right]\nonumber\\ \\ \bar\psi(x)&=&\sum_{\bf k}\left(\frac
{m}{k_{0}}\right)^{1/2}\left[b^{\dag}_{{\bf k},1}(t)\bar u_{1}({\bf
k}) \frac {e^{-i{\bf kx}}}{\sqrt {L}}+b_{{\bf k},2}(t)\bar u_{2}({\bf
k}) \frac {e^{i{\bf kx}}}{\sqrt {L}}\right] \;\;,\nonumber \\
\nonumber
\end{eqnarray}

\noindent where $b^{\dag}_{{\bf k},1}$ and $b_{{\bf k},1}$
[$b^{\dag}_{{\bf k},2}$ and $b_{{\bf k},2}$] are fermion creation and
annihilation operators associated with positive[negative]-energy
solutions $u_{1}({\bf k})$ [$u_{2}({\bf k})$] of Dirac's equation.
Canonical quantization demands that the creation and annihilation
operators satisfy the standard anticommutation relations at equal
times

\begin{eqnarray}
\{b^{\dag}_{{\bf k},\lambda}(t),b_{{\bf
k'},\lambda'}(t')\}_{t=t'}&=&\delta_{{\bf k},{\bf k}'}
\delta_{\lambda,\lambda'}\;\;\;\;{\rm for}\;\;\;\lambda,\lambda'=1,2
\nonumber\\ \\ \{b^{\dag}_{{\bf k},\lambda}(t),b^{\dag}_{{\bf
k'},\lambda'}(t')\}_{t=t'}&=&\{b_{{\bf k},\lambda}(t),b_{{\bf
k'},\lambda'}(t')\}_{t=t'}= 0 \;\;.\nonumber \\ \nonumber
\end{eqnarray}

\noindent In Eq.(1) ${\bf x}$ denotes the spatial coordinate only and
we use the notation

\[
(k_{0})^{2}=({\bf k})^{2}+m^{2}\;\;\;\; {\rm and}\;\;\;\;
kx=k_{0}t-{\bf k}{\bf x} \;.
\] 

In general, the state of the system is given in terms of a many-body
density operator ${\cal F}$ in the Heisenberg picture, a time
independent, non-negative, Hermitian operator with unit trace. At this
point, however, we specialize the analysis to the case of spatially
uniform systems. These systems exhibit translational invariance
(homogeneity) and rotational invariance (isotropy). In the case of
(1+1) dimensions this reduces to invariance under translations and
under reflection. The possible nonvanishing mean values of bilinear
forms of field operators are

\begin{eqnarray}
R_{{\bf k},\lambda';{\bf k},\lambda}(t)&=& Tr [(b^{\dag}_{{\bf
k},\lambda' }(t)b_{{\bf k},\lambda}(t)){\cal F}]\;\;\;\; {\rm
for}\;\;\;\; \lambda,\lambda'=1,2 \nonumber\\ \\ \Pi_{{\bf
k},\lambda';{\bf k},\lambda}(t)&=& Tr [(b_{-{\bf k},\lambda'}
(t)b_{{\bf k},\lambda}(t)){\cal F}]\;\;\;\; {\rm for}\;\;\;\;
\lambda,\lambda'=1,2 \;.  \nonumber
\end{eqnarray}
\vskip 0.5cm

\noindent The Hermitian matrix $R$ and the antisymmetric matrix $\Pi$
are the one-fermion density and pairing density respectively. Using
these objects we can construct the extended one-body density \cite{12}

\begin{equation}
{\cal R}_{\bf k}(t)=\left[
\begin{array}{cc}
   R_{\bf k}(t)       &  \Pi_{\bf k}(t)\\
                   &             \\
-\Pi^{*}_{\bf k}(t)   & I_{2}-R^{*}_{\bf k}(t)
\end{array}
\right]=
\left[
\begin{array}{cc}
\langle b^{\dag}_{{\bf k},\lambda'}(t)b_{{\bf k},\lambda}(t)\rangle &
\langle b_{-{\bf k},\lambda'}(t)b_{{\bf k},\lambda}(t)\rangle \\
                                 &                                  \\
\langle b^{\dag}_{-{\bf k},\lambda'}(t)b^{\dag}_{{\bf k},\lambda}(t)
\rangle &
\langle b_{{\bf k},\lambda'}(t)b^{\dag}_{{\bf k},\lambda}(t)\rangle
\end{array}
\right].
\end{equation}
\vskip 0.5cm

\noindent This object summarizes all information on the Gaussian
observables and provides an adequate starting point for our kinetic
treatment. 

The first step is standard and consists in reducing the extended
one-body density to diagonal form. This can be achieved by subjecting
the creation and annihilation operators to a canonical transformation
of the Bogolyubov type with coefficients defined by the eigenvalue
problem

\begin{equation}
{\cal X}^{\dag}_{\bf k}(t){\cal R}_{\bf k}(t) {\cal X}_{\bf k} (t) =
Q_{\bf k}(t)\;,
\end{equation}
\vskip 0.5cm

\noindent where the unitary matrix ${\cal X}_{\bf k}(t)$ which
diagonalizes ${\cal R}_{\bf k}(t)$ has the structure

\begin{equation}
{\cal X}_{\bf k}=\left[
\begin{array}{cc}
X^{*}_{\bf k}  &  Y^{*}_{\bf k}\\
               &               \\
  Y_{\bf k}    &    X_{\bf k}
\end{array}
\right]\; ; \;\;\;\;\;\;\;\;{\cal X}^{\dag}_{\bf k} =
\left[
\begin{array}{cc}
X^{T}_{\bf k}  &  Y^{\dag}_{\bf k}\\
               &                  \\
Y^{T}_{\bf k}  &  X^{\dag}_{\bf k}
\end{array}
\right]
\end{equation}
\vskip 0.5cm

\noindent and $Q_{\bf k}(t)$ is a diagonal matrix which can be written
as

\begin{equation}
Q_{\bf k}(t)=\left[
\begin{array}{cc}
\nu_{\bf k}(t)  &        0         \\
             &                  \\
      0      & I_{2}-\nu_{\bf k}(t)
\end{array}
\right]
=\left[
\begin{array}{cc}
\langle \beta^{\dag}_{{\bf k},\lambda'}(t)
\beta_{{\bf k},\lambda}(t)\rangle &
\langle \beta_{-{\bf k},\lambda'}(t)
\beta_{{\bf k},\lambda}(t)\rangle \\
                                &                                    \\
\langle \beta^{\dag}_{-{\bf k},\lambda'}(t)
\beta^{\dag}_{{\bf k},\lambda}(t)
\rangle &
\langle \beta_{{\bf k},\lambda'}(t)
\beta^{\dag}_{{\bf k},\lambda}(t)\rangle
\end{array}
\right]\;.
\end{equation}
\vskip 0.5cm

\noindent The unitary conditions for ${\cal X}_{\bf k}(t)$ can be
interpreted as orthogonality and completeness relations for a set of
natural orbitals which diagonalize the extended one-body density. They
read

\begin{equation}
{\cal X}_{\bf k}^{\dag}{\cal X}_{\bf k}=I_{4} \;\;\;{\rm and}\;\;\;
{\cal X}_{\bf k}{\cal X}^{\dag}_{\bf k}=I_{4}\;\;,
\end{equation}
\vskip 0.4cm

\noindent or, more explicitly, in terms of the submatrices $X_{\bf k}$
and $Y_{\bf k}$ [see Eq.(6)]

\vskip -0.2cm
\begin{eqnarray}
&&Y_{\bf k}Y^{\dag}_{\bf k}+X_{\bf k}X^{\dag}_{\bf k}=
{\bf I}_{2} \;\;\;\; , \;\;\;\; 
Y_{\bf k}X_{\bf k}^{T}+X_{\bf k}Y_{\bf k}^{T}={\bf 0}_{2} 
\;\;,\nonumber\\ \\
&&Y_{\bf k}^{\dag}Y_{\bf k}+X_{\bf k}^{T}X_{\bf k}^{*}=
{\bf I}_{2} \;\;\;\; , \;\;\;\; 
Y_{\bf k}^{T}X_{\bf k}^{*}+X_{\bf k}^{\dag}Y_{\bf k}=
{\bf 0}_{2}\;.\nonumber
\end{eqnarray}

\noindent The elements of the diagonal submatrix $\nu_{\bf k}(t)$ can,
on the other hand, be interpreted as quasi-fermion occupation numbers
for the paired natural orbitals. Because of the assumed symmetry one
must have

\vskip -0.2cm
\begin{equation}
\nu_{{\bf k},\lambda}(t)=\nu_{-{\bf k},\lambda}(t)\;.
\end{equation}
\vskip 0.4cm

\noindent Finally, using Eq.(6), we can relate the fermion operators
$b_{{\bf k},\lambda}^{\dag}(t)$ and $b_{{\bf k},\lambda}(t)$ to the
new quasi-fermion operators $\beta_{{\bf k},\lambda}^{\dag}(t)$ and
$\beta_{{\bf k},\lambda}(t)$ for $ \lambda=1,2$ as

\begin{equation}
\left[
\begin{array}{c}
    b_{{\bf k},1}     \\
                      \\ 
    b_{{\bf k},2}     \\
                      \\ 
b_{-{\bf k},1}^{\dag} \\
                      \\   
b_{-{\bf k},2}^{\dag}
\end{array}
\right]=\left[
\begin{array}{cccc}
X_{11}    &    X_{21}    &  Y_{11}^{*}  &  Y_{21}^{*}  \\
          &              &              &              \\  
X_{12}    &    X_{22}    &  Y_{12}^{*}  &  Y_{22}^{*}  \\
          &              &              &              \\ 
Y_{11}    &    Y_{21}    &  X_{11}^{*}  &  X_{21}^{*}  \\
          &              &              &              \\
Y_{12}    &    Y_{22}    &  X_{12}^{*}  &  X_{22}^{*}
\end{array}
\right]\left[
\begin{array}{c}
    \beta_{{\bf k},1}    \\
                         \\ 
    \beta_{{\bf k},2}    \\
                         \\
\beta_{-{\bf k},1}^{\dag}\\
                         \\
\beta_{-{\bf k},2}^{\dag}
\end{array}
\right]\;.
\end{equation}

\vskip 0.5cm

\noindent With the help of Eq.(11) it is then an easy task to express
$\bar\psi(x)$ and $\psi(x)$, Eq.(1), in term of $\beta_{{\bf
k},\lambda}^{\dag}(t)$ and $\beta_{{\bf k},\lambda}(t)$ for $
\lambda=1,2$. In doing so, one finds that the plane waves of
$\bar\psi(x)$ and $\psi(x)$ are modified by a complex,
momentum-dependent redefinition of $m$ involving the Bogolyubov
parameters. The complex character of these parameters is actually
crucial in dynamical situations, where the imaginary parts will allow
for the description of time-odd (i.e., velocity-like) properties.

What we have achieved so far amounts to obtaining an expansion of the
fields $\bar\psi(x)$ and $\psi(x)$ such that the mean values in ${\cal
F}$ of Gaussian observables are parametrized in terms of the natural
orbitals $X_{\lambda^{\prime},\lambda}(\bf k)$ and
$Y_{\lambda^{\prime},\lambda}(\bf k)$ and of the occupation numbers
$\nu_{{\bf k},\lambda}(t)= Tr \left(\beta^{\dag}_{{\bf
k},\lambda}\beta_{{\bf k},\lambda} {\cal F}\right)$ for $
\lambda=1,2$.  In general, all these quantities are time dependent
under the Heisenberg dynamics of the field operators, and we now
proceed to obtain the corresponding equations of motion.  Taking the
time derivative of Eq.(5) and using the unitarity condition (8) we get

\begin{equation}
{\cal X}^{\dag}_{\bf k}\dot {\cal R}_{\bf k}
{\cal X}_{\bf k}=\dot Q_{\bf k}-\dot 
{\cal X}^{\dag}_{\bf k}{\cal X}_{\bf k}Q_{\bf k}-
Q_{\bf k}{\cal X}^{\dag}_{\bf k}
\dot {\cal X}_{\bf k}\;.
\end{equation}
                                    
\vskip 0.5cm

\noindent We next evaluate the left-hand side of this equation using
the Heisenberg equation of motion to obtain

\begin{equation}
i{\cal X}^{\dag}_{\bf k}\dot {\cal R}_{\bf k}{\cal X}_{\bf k}=
\left[
\begin{array}{cc}
Tr\left(\left[\beta_{{\bf k},\lambda^{\prime}}^{\dag}
\beta_{{\bf k},\lambda},H\right]
{\cal F}\right)  &
Tr\left(\left[\beta_{-{\bf k},\lambda^{\prime}}
\beta_{{\bf k},\lambda},H\right]
{\cal F}\right)  \\
                                   &                      \\
Tr\left(\left[\beta_{-{\bf k},\lambda^{\prime}}^{\dag}
\beta_{{\bf k},\lambda}^{\dag},H\right]
{\cal F}\right)  &
Tr\left(\left[\beta_{{\bf k},\lambda^{\prime}}
\beta_{{\bf k},\lambda}^{\dag},H\right]
{\cal F}\right)  \\
\end{array}
\right]\;.
\end{equation}

\vskip 0.5cm

\noindent The right-hand side of Eq.(12) can also be evaluated
explicitly using Eqs.(6) and (7):

\begin{equation}
i \left(\dot Q_{\bf k}-
\dot {\cal X}^{\dag}_{\bf k}{\cal X}_{\bf k}Q_{\bf k}-
Q_{\bf k}{\cal X}^{\dag}_{\bf k}\dot {\cal X}_{\bf k}\right)=\left[
\begin{array}{cc}
i \dot \nu_{\bf k}+[\nu_{\bf k},h^{*}_{\bf k}]_{-}  &  
-g^{*}_{\bf k}+\{\nu_{\bf k},g^{*}_{\bf k}\}_{+} \\
                                              &
                                                  \\
-g_{\bf k}+\{\nu_{\bf k},g_{\bf k}\}_{+}  &
-i \dot \nu_{\bf k}+[\nu_{\bf k},h_{\bf k}]_{-}
\end{array}
\right]\;,
\end{equation}

\vskip 0.5cm
\noindent
where the matrices $h_{\bf k}$ and $g_{\bf k}$ are given in terms of
$X_{\bf k}$ and $Y_{\bf k}$ as

\begin{eqnarray}
h_{\bf k}&=&-i(\dot Y^{T}_{\bf k}Y^{*}_{\bf k}+
\dot X^{\dag}_{\bf k}X_{\bf k}) \nonumber\\ \\
g_{\bf k}&=&-i(\dot Y^{T}_{\bf k}X^{*}_{\bf k}+
\dot X^{\dag}_{\bf k}Y_{\bf k}) \;.\nonumber 
\end{eqnarray}

\vskip 0.5cm

\noindent From Eqs.(13) and (14) we obtain dynamical equations which
describe the time evolution of our uniform fermion system. They read

\begin{eqnarray}
i \dot \nu_{\bf k}+[\nu_{\bf k},h^{*}_{\bf k}]_{-} &=& 
Tr \left([\beta^{\dag}_{{\bf k},\lambda'}\beta_{{\bf k},\lambda},H]
{\cal F}\right) \nonumber\\ \\
-g^{*}_{\bf k}+\{\nu_{\bf k},g^{*}_{\bf k}\}_{+} &=& 
Tr \left([\beta_{-{\bf k},\lambda'}\beta_{{\bf k},\lambda},H]
{\cal F}\right) \;.\nonumber\\ \nonumber
\end{eqnarray}

Eqs.(16), together with the unitarity conditions (9), determine the
time rate of change of the Gaussian observables in terms of
expectation values of appropriate commutators. They are, however,
clearly not closed equations when the Hamiltonian $H$ involves
self-interacting fields. In this case, in fact, the time derivatives
of the Gaussian observables are given in terms of traces which are not
expressible in terms of the Gaussian observables themselves, since
they will involve also many-fermion densities. This situation can be
dealt with in terms of the projection technique reviewed in the next
section.

\section{Projection technique and approximation scheme}

In this section we introduce the time-dependent projection technique
\cite{7} which permits to obtain closed approximations to the
equations of motion (16).  It has been developed earlier in the
context of nonrelativistic nuclear many-body dynamics and was recently
applied in the quantum-field theoretical context to the
self-interacting $\lambda \phi^{4}$ theory in (1+1) dimensions
\cite{9}. It allows for the formulation of a mean-field expansion for
the dynamics of the two-point correlation function from which one
recovers the results of the Gaussian mean-field approximations in
lowest order. If carried to higher orders it allows for the inclusion
and evaluation of higher dynamical correlation corrections to the
simplest mean-field approximation.

In the specific context of the equations of motion (16) we begin by 
decomposing the full density ${\cal F}$ as 

\begin{equation}
{\cal F}={\cal F}_{0}(t)+{\cal F}'(t)\;\;,
\end{equation}

\vskip 0.5cm

\noindent where ${\cal F}_{0}(t)$ is a Gaussian ansatz which achieves
a Hartree-Fock factorization of traces involving more than two field
operators. The Gaussian density ${\cal F}_{0}(t)$ is chosen as having
the form of a exponential of a bilinear, Hermitian expression in the
fields normalized to unit trace \cite{5}. In the momentum basis, it
reads

\vskip 0.5cm
\begin{equation}
{\cal F}_{0}=\frac{\exp \left[\sum_{({\bf k}_{1},{\bf k}_{2})}
A_{{\bf k}_{1},{\bf k}_{2}}b^{\dag}_{{\bf k}_{1}}b_{{\bf k}_{2}}+
B_{{\bf k}_{1},{\bf k}_{2}}b^{\dag}_{{\bf k}_{1}}b^{\dag}_{{\bf k}_{2}}+
C_{{\bf k}_{1},{\bf k}_{2}}b_{{\bf k}_{1}}b_{{\bf k}_{2}}\right]}
{Tr \left\{\exp \left[\sum_{({\bf k}_{1},{\bf k}_{2})}
A_{{\bf k}_{1},{\bf k}_{2}}b^{\dag}_{{\bf k}_{1}}b_{{\bf k}_{2}}+
B_{{\bf k}_{1},{\bf k}_{2}}b^{\dag}_{{\bf k}_{1}}b^{\dag}_{{\bf k}_{2}}+
C_{{\bf k}_{1},{\bf k}_{2}}b_{{\bf k}_{1}}b_{{\bf k}_{2}}\right]\right\}}\;.
\end{equation}

\vskip 0.5cm

\noindent The parameters in Eq.(18) are fixed by requiring that mean
values in ${\cal F}_{0}$ of expressions that are bilinear in the
fields reproduce the corresponding ${\cal F}$ averages [see Eqs.(20)
below]. ${\cal F}_{0}$ is a time-dependent object, which acquires a
particularly simple form when expressed in terms of the Bogolyubov
quasi-fermion operators

\begin{equation}
{\cal F}_{0}(t)=\prod_{{\bf k},\lambda}\left[ \nu_{{\bf k},\lambda}
\beta_{{\bf k},\lambda}^{\dag}(t)\beta_{{\bf k},\lambda}(t)+
(1-\nu_{{\bf k},\lambda})\beta_{{\bf k},\lambda}(t)
\beta_{{\bf k},\lambda}^{\dag}(t)\right]\;.
\end{equation}

\vskip 0.5cm
\noindent This is clearly a unit trace object which in addition
satisfies

\begin{eqnarray}
&&Tr\left(\beta_{a}{\cal F}_{0}\right)=Tr\left(\beta_{a}{\cal F}\right)=
Tr\left(\beta_{a}^{\dag}{\cal F}_{0}\right)=
Tr\left(\beta_{a}^{\dag}{\cal F}\right)=0 \;\;;\nonumber\\
&&Tr\left(\beta_{a}\beta_{b}{\cal F}_{0}\right)=
Tr\left(\beta_{a}\beta_{b}{\cal F}\right)=0 \;\;;\nonumber\\
&&Tr\left(\beta_{a}^{\dag}\beta_{b}^{\dag}{\cal F}_{0}\right)=
Tr\left(\beta_{a}^{\dag}\beta_{b}^{\dag}{\cal F}\right)=0 \;\;;\\
&&Tr\left(\beta_{a}^{\dag}\beta_{b}{\cal F}_{0}\right)=
Tr\left(\beta_{a}^{\dag}\beta_{b}{\cal F}\right)=
\nu_{a}\delta_{a,b} \;\;{\rm and} \nonumber\\
&&Tr\left(\beta_{a}\beta_{b}^{\dag}{\cal F}_{0}\right)=
Tr\left(\beta_{a}\beta_{b}^{\dag}{\cal F}\right)=
(1-\nu_{a})\delta_{a,b}\;. \nonumber\\ \nonumber
\end{eqnarray}

\noindent Correspondingly, the ``remainder'' density ${\cal
F}^{\prime}(t)$, defined by Eq.(17), is a traceless, pure correlation
density. As already remarked, a crucial point to observe is that
${\cal F}_{0}(t)$ can be written as a time-dependent projection of
${\cal F}$, i.e.,

\begin{eqnarray}
{\cal F}_{0}(t) = {\cal P}(t){\cal F}\;\;\;{\rm with}\;\;\;
{\cal P}(t){\cal P}(t)={\cal P}(t)\;.
\end{eqnarray}
\vskip 0.3cm

\noindent In order to completely define this projector we require
further that it satisfies

\begin{equation}
i \dot {\cal F}_{0}(t)=\left[{\cal P}(t),{\cal L}\right]{\cal F}=
\left[{\cal F}_{0}(t),H\right]+{\cal P}(t)\left[H,{\cal F}\right]\;,
\end{equation}

\vskip 0.5cm

\noindent where ${\cal L}$ is the Liouvillian defined as

\begin{equation}
{\cal L}\;\cdot=[H,\;\cdot\;]\;,
\end{equation}

\vskip 0.5cm

\noindent $H$ being the Hamiltonian of the field. Eq.(22) is just the
Heisenberg picture counterpart of the condition $ \dot{{\cal P}}(t)
{\cal F} = 0$ which has been used to define ${\cal P}(t)$ in the
Schr\"odinger picture \cite{8}.  It is possible to show that
conditions (21) and (22) make ${\cal P}(t)$ unique and to obtain an
explicit form for this object in terms of the quasi-fermion operators
and of the natural orbital occupations \cite{3,7,8,9}. 

The existence of the projector ${\cal P}(t)$ allows one to obtain an
equation relating the correlation part ${\cal F}^{\prime}(t)$ to the
Gaussian part ${\cal F}_0(t)$ of the full density. This can be
immediately obtained from Eqs.(17), (21) and (22) and reads

\begin{equation}
\left(i\partial_{t}-{\cal P}(t){\cal L}\right){\cal F}'(t)=\left({\cal I}-
{\cal P}(t)\right){\cal L}{\cal F}_{0}(t)\;.
\end{equation}

\vskip 0.5cm

\noindent This equation has the formal solution

\begin{equation}
{\cal F}'(t)={\cal G}(t,0){\cal F}'(0)-i\int_{0}^{t} dt'\;{\cal
G}(t,t') \left({\cal I}-{\cal P}(t')\right){\cal L}{\cal
F}_{0}(t')\;,
\end{equation}

\vskip 0.5cm

\noindent where the first term accounts for initial correlations
possibly contained in ${\cal F}$. The object ${\cal G}(t,t^{\prime})$
is the time-ordered Green's function

\begin{equation}
{\cal G}(t,t')= 
T\left(\exp{\left[i\int_{t'}^{t}d
\tau{\cal P}(\tau){\cal L}\right]}\right)\;.
\end{equation}

\vskip 0.5cm

We see thus that ${\cal F}^{\prime}(t)$, and therefore also ${\cal F}$
[see Eq.(17)], can be formally expressed in terms of ${\cal
F}_{0}(t^{\prime})$ (for $t^{\prime}\leq t$) and of initial
correlations ${\cal F}^{\prime}(0)$. This allows us to express also
the dynamical equations (16) as functionals of ${\cal
F}_{0}(t^{\prime})$ and of the initial correlations.  Since, on the
other hand, the reduced density ${\cal F}_{0}(t^{\prime})$ is
expressed in terms of the one-fermion densities alone, we see that the
resulting equations are now essentially closed equations. Note,
however, that the complicated time dependence of the field operators
is explicitly probed through the memory effects present in the
expression (25) for ${\cal F}^{\prime}(t)$. Approximations are
therefore needed for the actual evaluation of this object.  A
systematic expansion scheme for the memory effects has been discussed
in Refs. \cite{3,8,9}.  The lowest order correlation corrections to the
pure mean field approximation, in which ${\cal F}'$ is simply ignored,
correspond to replacing the full Heisenberg time-evolution of
operators occurring in the collision integrals by a mean-field
evolution governed by 

\[
H_{0}={\cal P}^{\dag}(t)H\;\;.\nonumber
\]

\vskip 0.4cm

\noindent Consistently with this approximation, ${\cal L}$ is replaced
in (25) and (26) by ${\cal L}_{0}\;\cdot=[H_{0},\;\cdot\;]$.  In this
way correlation effects are treated to second order in $H$ in the
resulting collision integrals.

An important feature of this scheme (which holds also for higher
orders of the expansion \cite{8}) is that the mean energy is
conserved, namely

\[
\frac{\partial}{\partial t}\langle H\rangle=0
\]

\noindent where

\[
\langle H\rangle=Tr\;H{\cal F}_{0}(t)+Tr\;H{\cal F}^{\prime}(t)\;.
\]

In the following sections we apply the general expressions obtained in
above to treat a uniform fermion system described by Chiral
Gross-Neveu model (CGNM). We will consider only the lowest
(mean-field) approximation, corresponding to ${\cal F}^{\prime}(t)=0$.
Collisional correlations will be treated elsewhere.

\section{The Chiral Gross-Neveu model (CGNM)}

The Hamiltonian density for the CGNM is given by

\begin{equation}
{\cal H}_{\mbox{\tiny CGNM}}=\sum_{i=1}^{N}\left\{\bar\psi^{i}
\left[-i\gamma_{1}\partial_{1}\right]\psi^{i}\right\}- \frac
{g^{2}}{2}\left\{\left[\sum_{i=1}^{N}\bar\psi^{i}\psi^{i}\right]^{2}-
\xi\left[\sum_{i=1}^{N}\bar\psi^{i}\gamma_{5}\psi^{i}
\right]^{2}\right\}\;,
\end{equation}

\vskip 0.7cm

\noindent where $\xi$ is a constant which indicates whether the model
is invariant under the discrete $\gamma_{5}$ transformation ($\xi=0$)
or under the Abelian chiral $U(1)$ group ($\xi=1$).

This is a massless fermion theory in (1+1) dimensions with quartic
interaction. The model contains $N$ species of fermions coupled
symmetrically, where $\psi^{i}$ is a complex Dirac spinor transforming
as the fundamental representation of $SU(N)$ group. It is known that
the actual symmetry of the theory is not $SU(N)$ but rather $O(2N)$
\cite{DHN}.  The transformations forming this group mix not only
particles but also particles with antiparticles.  This model is
essentially equivalent to the Nambu-Jona-Lasinio model \cite{NJ},
except for the fact that in (1+1) dimensions it is renormalizable.
Moreover, it is one of the very few known field theories which are
assimptotically free.  To leading order in a $1/N$ expansion \cite{10},
the CGNM exhibits a number of interesting phenomena, like spontaneous
symmetry breaking \cite{EW}, dynamical fermion mass generation and
dimensional transmutation.  The model possess an infinite number of
conservation laws, and as a consequence, the $S$-matrix may be
computed exactly \cite{BKKW}.

To obtain the time evolution of Bogolyubov parameters we have to
obtain the CGNM Hamiltonian [see Eq.(16)] by integrating over
one-dimensional space. This involves, in particular, choosing a
representation for the $\gamma$-matrices.  Here we have to be careful,
since a bad choice of representation can spoil manifest reflection
invariance. In Appendix A (see also Ref.\cite{NJ}) we discuss suitable
representations for the $\gamma$-matrices. We choose the Pauli-Dirac
representation, namely

\begin{equation}
\gamma_{0}=\sigma_{3}\;\;\;;\;\;\;\gamma_{1}=i\sigma_{2}\;\;\;
{\rm and}\;\;\; \gamma_{5}=\gamma_{0}\gamma_{1}=\sigma_{1}\;.
\end{equation}

\vskip 0.5cm 

\noindent In this representation the spinors $u_{1}({\bf k})$ and
$u_{2}({\bf k})$ are given by

\begin{equation}
u_{1}({\bf k})=\left[\frac {(k_{0}+m)}{2m}\right]^{1/2}\left[
\begin{array}{c}
           1           \\
                       \\  
        {\bf k}        \\
\overline{(k_{0}+m)}
\end{array}
\right]\;\;\;,\;\;\; u_{2}({\bf k})=
\left[\frac {(k_{0}-m)}{2m}\right]^{1/2}\left[
\begin{array}{c}
           1           \\
                       \\
        {\bf k}        \\
\overline{(k_{0}-m)}
\end{array}
\right]\;.
\end{equation}

\vskip 0.5cm 

\noindent The resulting form for the CGNM Hamiltonian obtained using
this representation is given in full in Appendix B.

We next consider the initial conditions problem for this system.  The
general Bogolyubov transformation defined in (11) breaks both chiral
and charge symmetries, but we restrict the following development to a
special Bogolyubov transformation (to be called Nambu transformation)
which breaks the chiral symmetry of our system only.  The elements of
this Nambu transformation, parametrized consistently with the
unitarity conditions (9), are given by

\begin{eqnarray}
&&X_{11}=X_{22}=\cos\varphi_{\bf k}\;\;\;
{\rm and}\;\;\;X_{12}=X_{21}=0 \nonumber\\ \\
&&Y_{12}=-Y_{21}=\sin\varphi_{\bf k}e^{i\gamma_{\bf k}}\;\;\;
{\rm and}\;\;\;Y_{11}=Y_{22}=0\;.
\nonumber
\end{eqnarray} 

\vskip 0.4cm

\noindent In the special case of a Nambu transformation, the elements
of the matrices $h_{\bf k}$ and $g_{\bf k}$ [see Eq.(15)] are given by

\begin{eqnarray}
&&h_{11}=h_{22}=\dot\gamma_{\bf k}\sin^{2}\varphi_{\bf k}
\nonumber\\ 
&&h_{12}=h_{21}=g_{11}=g_{22}=0
\\ 
&&g_{12}=-g_{21}=[i\dot\varphi_{\bf k}-\dot \gamma_{\bf k}
\sin\varphi_{\bf k}\cos\varphi_{\bf k}]e^{i\gamma_{\bf k}}\;.\nonumber 
\end{eqnarray}

\vskip 0.5cm

\noindent On the other hand, in the mean-field approximation one finds
that ${\cal F}_{0}(t)$ commutes with number operators

\begin{equation}
Tr \;\{H_{\mbox{\tiny CGNM}}[{\cal F}_{0},\beta_{{\bf k},1}^{\dag}
\beta_{{\bf k},1}]\}= Tr\; \{H_{\mbox{\tiny CGNM}}
[{\cal F}_{0},\beta_{{\bf k},2}^{\dag}
\beta_{{\bf k},2}]\}=0\;,
\end{equation}

\vskip 0.4cm

\noindent while, due to the charge conservation in the system, we
have also

\begin{eqnarray}
&&Tr \;\{[\beta_{{\bf k},1}^{\dag}
\beta_{{\bf k},2},H_{\mbox{\tiny CGNM}}]{\cal F}_{0}\}=
Tr \;\{[\beta_{{\bf k},2}^{\dag}
\beta_{{\bf k},1},H_{\mbox{\tiny CGNM}}]{\cal F}_{0}\}=0
\nonumber\\ \\
&&Tr \;\{[\beta_{-{\bf k},1}\beta_{{\bf k},1},
H_{\mbox{\tiny CGNM}}]{\cal F}_{0}\}= 
Tr \;\{[\beta_{-{\bf k},2}\beta_{{\bf k},2},
H_{\mbox{\tiny CGNM}}]{\cal F}_{0}\}=0\;.
\nonumber\\ \nonumber 
\end{eqnarray}

Substituting Eqs. (31), (32) and (33) in Eq. (16), we obtain the
equations which describe the time evolution of our system as

\begin{equation}
\dot\nu_{{\bf k},1}=0\;\;\;{\rm and}\;\;\;\dot\nu_{{\bf k},2}=0
\end{equation}

\begin{equation}
\left[i\dot\varphi_{\bf k}+\dot\gamma_{\bf k}
\sin\varphi_{\bf k}\cos\varphi_{\bf k}\right]
e^{-i\gamma_{\bf k}}=\frac {Tr\left([
\beta_{-{\bf k},1}\beta_{{\bf k},2},
H_{\mbox{\tiny CGNM}}]{\cal F}_{0}\right)}{(1-\nu_{{\bf k},1}-
\nu_{{\bf k},2})}\;.
\end{equation}

\vskip 0.7cm

Equation (34) shows that the occupation numbers of the paired natural
orbitals are constant, i.e., we recover the general isoentropic
character of the mean-field approximation. The complex equation of
motion (35) describes the time evolution of the Nambu
parameters. Writing the CGNM Hamiltonian, given in Appendix B, in the
Nambu basis using Eqs.(11) and (30), and substituting this Hamiltonian
in Eq.(35), we obtain the explicit form of these equation. The
calculation of traces is lengthy but straightforward. As a result one
obtains

\begin{eqnarray}
i\dot\varphi_{\bf k}&+&\dot\gamma_{\bf k}\frac {\sin 2\varphi_{\bf
k}}{2}= \frac{({\bf k})^{2}}{k_{0}}\sin 2\varphi_{\bf k}-m\frac {|{\bf
k}|}{k_{0}} \left[\sin^{2}\varphi_{\bf k}e^{-i{\gamma_{\bf
k}}}-\cos^{2} \varphi_{\bf k}e^{i\gamma_{\bf k}}\right]+\nonumber\\ \\
&+&\left(\frac {g^{2}m^{2}}{4\pi}\right)\frac{(\xi+1)}{k_{0}}
\left[\sin 2\varphi_{\bf k}+\frac {|{\bf
k}|}{m}\left(\sin^{2}\varphi_{\bf k} e^{-i\gamma_{\bf
k}}-\cos^{2}\varphi_{\bf k}e^{i\gamma_{\bf k}}\right)\right]
(I_{1}+I_{2})\nonumber\\ \nonumber
\end{eqnarray}

\noindent where $I_{1}$ and $I_{2}$ are the divergent integrals 

\begin{eqnarray}
I_{1}&=&\int \frac {d{\bf k}'}{k_{0}'}\cos 2\varphi_{{\bf k}'}
(1-\nu_{{\bf k}',1}-\nu_{{\bf k}',2})\nonumber\\ \\
I_{2}&=&\int \frac {d{\bf k}'}{k_{0}'}\frac {|{\bf k}'|}{m}
\sin 2\varphi_{{\bf k}'}\cos\gamma_{{\bf k}'}(1-\nu_{{\bf k}',1}-
\nu_{{\bf k}',2})\;.\nonumber\\ \nonumber
\end{eqnarray}

\noindent We take $N=1$ for simplicity. Splitting the complex equation
(36) into real and imaginary parts we have

\begin{eqnarray}
\dot\varphi_{\bf k}&=&\sin\gamma_{\bf k}\frac {|{\bf
k}|}{k_{0}}\left[m- \left(\frac {g^{2}m}{4\pi}\right)(\xi+1)
(I_{1}+I_{2})\right]\nonumber\\ \nonumber\\ \dot\gamma_{\bf k}\sin
2\varphi_{\bf k}&=& \frac {2\sin 2\varphi_{\bf k}}{k_{0}} \left[{\bf
k}^{2}+\left(\frac {g^{2}m^{2}}{4\pi}\right)
(\xi+1)(I_{1}+I_{2})\right]+ \\ \nonumber\\ &+&2\cos 2\varphi_{\bf
k}\cos\gamma_{\bf k}\frac {|{\bf k}|}{k_{0}} \left[m-\left(\frac
{g^{2}m}{4\pi}\right)(\xi+1)(I_{1}+I_{2})\right] \;.\nonumber\\
\nonumber
\end{eqnarray}

Finally, the mean-field energy is evaluated as

\begin{eqnarray}
\langle H_{\mbox{\tiny CGNM}}^{\mbox{\tiny M.F.}}\rangle &=&
Tr\left[H_{\mbox{\tiny CGNM}}{\cal F}_{0}(t)\right]= \left(\frac
{m^2}{2\pi}\right)(I_{2}-I_{3})- \left(\frac
{g^{2}m^{2}}{8\pi^{2}}\right) \frac {(\xi+1)}{2}(I_{1}+I_{2})^{2}\;+
\nonumber\\ \\ &-&\left(\frac {g^{2}}{8\pi^{2}}\right)\frac
{(\xi+1)}{2}\int d{\bf k}' (1+\nu_{{\bf k}',1}-\nu_{{\bf k}',2})\int
d{\bf k}''(1-\nu_{{\bf k}'',1}+ \nu_{{\bf k}'',2})\;\;\;, \nonumber\\
\nonumber
\end{eqnarray}

\noindent where $I_{1}$ and $I_{2}$ are given in (37) and $I_{3}$ is
given by

\begin{equation}
I_{3}=\int \frac {d{\bf k}'}{k_{0}'}\left(\frac {{\bf k}'}{m}\right)^{2}
\cos 2\varphi_{\bf k}(1-\nu_{{\bf k}',1}-\nu_{{\bf k}',2})\;\;.
\end{equation}

\vskip 0.4cm

Since all the results above contain divergent integrals,
a renormalization procedure is required. This is discussed in the next
section.

\section{Renormalization}

In order to handle the infinities which occur in the preceding
equations it is necessary to introduce a renormalization prescription
that will render physical quantities finite. In general,
renormalization procedures consist in combining divergent terms with
the bare mass and coupling constants of the theory to define new,
finite (or renormalized) values of these quantities. Thus, when one
adopts a momentum cut-off to regularize divergent integrals, the bare
mass and coupling constants are chosen to be cut-off dependent in a
way that will cancel the divergences. In the present case, however the
divergent integrals (37) and (40) involve the dynamical variables
themselves in the integrand, so that even their degree of divergence
is not directly computable. In order to handle this situation we will
use a self-consistent renormalization procedure inspired in
Ref.\cite{TOM}.

The renormalization prescription we use can be based on the
consideration of the static solutions of the dynamical equations
(38). They are determined by the equations

\begin{eqnarray}
&& \sin\gamma_{\bf k}|_{\mbox{\scriptsize eq}} \left[1-\left(\frac
{g^{2}}{4\pi}\right) (\xi+1)(I_{1}+I_{2})\right]= 0 \\ \nonumber\\ &&
\tan 2\varphi_{\bf k}|_{\rm eq}= \frac{-|{\bf
k}|m\left[1-\left(g^{2}/4\pi\right) (\xi+1)(I_{1}+I_{2})\right]}
{\left[({\bf k})^{2}+\left(g^{2}m^{2}/4\pi\right)
(\xi+1)(I_{1}+I_{2})\right]} \cos\gamma_{\bf k}|_{\rm eq}\;.\\
\nonumber
\end{eqnarray}

\noindent We will then show explicitly that it also controls the
divergences which appear in the kinetic regime of the mean-field
approximation.

In order to obtain the renormalization prescription we introduce a
momentum cut-off $\Lambda$ and begin by assuming that, in order to
render the theory finite, the bare coupling constant $g^2$ must
approach zero for large values of $\Lambda$ as (see
e.g. Refs. \cite{10}, \cite{RK})

\begin{equation}
g^{2}=\frac{4\pi}{(\xi+1)}\left[\ln\left(\frac {\Lambda^{2}}{m^{2}}
\right) \right]^{-1}\; ,
\end{equation}

\vskip 0.5cm

\noindent where the form of the first factor is dictated by later
convenience. We next {\it assume} that the integrals $I_{1}$ and
$I_{2}$ have logarithmic divergences

\begin{eqnarray}
I_{1}&=& a+b\ln\left(\frac {\Lambda^{2}}{m^{2}}\right) \\
I_{2}&=& c+d\ln\left(\frac {\Lambda^{2}}{m^{2}}\right)\;,
\end{eqnarray}

\vskip 0.3cm

\noindent where $a$, $b$, $c$ and $d$ are finite constants.
Substituting (43) and the ansatze (44) and (45) in the static equation
(42) we obtain

\begin{equation}
\tan 2\varphi_{\bf k}|_{\rm eq}=\frac {-(-1)^{n}m|{\bf k}|
[1-(b+d)]}{[({\bf k})^{2}+m^{2}(b+d)]}\;\;.
\end{equation}

\vskip 0.5cm

\noindent Finally, evaluating the integrals $I_1$ and $I_2$,
Eqs.(37), using Eq.(46), we are able to determine the constants $a$,
$b$, $c$ and $d$ self-consistently. This calculation is given in
Appendix C. We find there $b=1$ while $d$ remains arbitrary. The
renormalized static equations are then obtained by simply substituting
these values in (46). We obtain

\begin{equation}
\tan 2\varphi_{\bf k}|_{\rm eq}=\frac {(-1)^{n}m|{\bf k}|d}{[{\bf k}^{2}+
(1+d)m^{2}]}\;.
\end{equation}

\vskip 0.3cm

\noindent We observe that the theory involves just one free parameter,
$d$. This is altogether reasonable since our starting point was a
massless fermion theory which was determined by one dimensionless
coupling constant $g^2$, and we end up with a theory determined by one
free parameter $d$ after the self-consistent renormalization
procedure. What is needed is an interpretation of the parameter $d$.

To obtain this, we begin by writing the fields $\psi({\bf x})$ and
$\bar\psi({\bf x})$ given in (1) in the Nambu quasi-particle basis
using (11), and (30). We find that the new Dirac spinors in this basis
are given by

\begin{eqnarray}
u_{1}'({\bf k})&=&\cos\varphi_{\bf k}\;u_{1}({\bf k})+
\sin\varphi_{\bf k}\;e^{i\gamma_{\bf k}}\;u_{2}(-{\bf k})\nonumber \\ \\
u_{2}'({\bf k})&=&\cos\varphi_{\bf k}\;u_{2}({\bf k})-
\sin\varphi_{\bf k}\;e^{-i\gamma_{\bf k}}\;u_{1}(-{\bf k})\;.\nonumber
\end{eqnarray}

\vskip 0.3cm

\noindent From the renormalized static solution (47) we have that 

\begin{eqnarray}
\cos\varphi_{\bf k}|_{\rm eq}&=&\frac {1}{2\sqrt{k_{0}x}}\left[(k_{0}x+
m|{\bf k}|d)^{1/2}+(k_{0}x-m|{\bf k}|d)^{1/2}\right]\nonumber\\ \\
\sin\varphi_{\bf k}|_{\rm eq}&=&(-1)^{n}\frac {1}{2\sqrt{k_{0}x}}
\left[(k_{0}x+m|{\bf k}|d)^{1/2}-(k_{0}x-m|{\bf k}|d)^{1/2}\right]\nonumber
\end{eqnarray}

\vskip 0.4cm

\noindent where $x=[k^{2}+(1+d)^{2}m^{2}]^{1/2}$.  Substituting the
solutions (49) and the particle spinors (29) in (48), we obtain the
new spinors in quasi-particle basis. They are given as

\begin{eqnarray}
u_{1}'({\bf k})&=&\left[
\frac{(k_{0}^{eff}+m_{eff})}{2m_{eff}}\right]^{1/2}\left[
\begin{array}{c}
      1     \\
            \\
   {\bf k}  \\
\overline{(k_{0}^{eff}+m_{eff})}
\end{array}
\right]\;, \nonumber\\ \\ 
u_{2}'({\bf k})&=&
\left[\frac {(k_{0}^{eff}-m_{eff})}{2m_{eff}}\right]^{1/2}\left[
\begin{array}{c}
      1     \\
            \\
   {\bf k}  \\
\overline{(k_{0}^{eff}-m_{eff})}
\end{array}
\right]\nonumber
\end{eqnarray}

\vskip 0.5cm

\noindent where 

\vskip -0.8cm
\begin{eqnarray}
(k_{0}^{eff})^{2}=({\bf k})^{2}+(m_{eff})^{2}\;\;\;
{\rm and}\;\;\;m_{eff}=(1+d)m\;.
\end{eqnarray}

\vskip 0.5cm

\noindent Comparing the spinors (50) in quasi-particle basis with the
spinors (29) in particle basis, we see in Eq.(51) what amounts to a
redefinition of the mass scale.  Therefore, we note that, unlike the
situation found in connection with the $1/N$ expansion, the use of the
Gaussian ansatz, Eq.(19), parametrized by the canonical transformation
leading to the quasi-fermion basis, allows for the direct dynamical
determination of the stable equilibrium situation of the system [see
Eqs.(41) and (42)], including symmetry breaking and mass generation.
Moreover, the renormalization procedure effectively replaces the
dimensionless coupling constant $g^2$ by the free parameter $d$
associated to the mass scale [see Eq.(51)]. This is analogous to the
phenomenon of dimensional transmutation found by Gross and Neveu
\cite{10} in the $1/N$ expansion. Finally, aside from the over-all
mass scale (characterized by $d$) there are no free adjustable
parameters.

Using Eqs.(38) and (43)-(45), we finally write the renormalized form
of the dynamical equations that describe the mean-field time evolution
of this system. As mentioned before they are now also finite and read

\begin{eqnarray}
\dot\nu_{{\bf k},1}= \dot\nu_{{\bf k},2}&=& 0 \nonumber\\ \nonumber\\
\dot\varphi_{\bf k}&=&-md\frac {|{\bf k}|}{k_{0}} \sin\gamma_{\bf
k}\nonumber\\ \\ \dot\gamma_{\bf k}\sin 2\varphi_{\bf k}&=& \frac
{2\sin 2\varphi_{\bf k}}{k_{0}} \left[({\bf
k})^{2}+m^{2}(1+d)\right]+\nonumber\\ \nonumber\\ &-&2md\frac{|{\bf
k}|}{k_{0}}\cos 2\varphi_{\bf k} \cos\gamma_{\bf k}\;.\nonumber
\end{eqnarray}
\vskip 0.3cm

\noindent The first two equations express just the isoentropic
character of the mean-field kinetics. The remaining coupled equations
describe the kinetic behavior of the Nambu parameters and are
discussed in the following section.

We conclude this section calculating the ground-state (vacuum)
energy of our system in the mean-field approximation. This
approximation to the ground state is obtained as a static solution of
Eqs.(52). The associated mean energy can be obtained by
taking $\nu_{{\bf k},1}=\nu_{{\bf k},2}=0$ in (39) and evaluating the
divergent integrals (37) and (40) using the renormalization ansatze
(43) and (44)-(45). We obtain

\begin{eqnarray}
\frac{\langle H_{\mbox{\tiny CGNM}}^{\mbox{\tiny M.F.}}\rangle^
{\mbox{\scriptsize vacuum}}}{L}&=&-\left(\frac{1}{2\pi}\right)
\left[1+\left(\frac{g^{2}}{2\pi}\right)(\xi+1)\right]\Lambda^{2}
\nonumber\\\nonumber\\
&-&\left(\frac{1}{2\pi}\right)\frac{(1+d)^{2}m^{2}}{2}
\left\{1-\ln\left[\frac{(1+d)^{2}}{4}\right]\right\}\;, \\ \nonumber
\end{eqnarray}

\noindent where we used for the finite constants $a$, $b$ and $c$ the
expresions given in Appendix C. The first term, divergent as
$\Lambda\rightarrow\infty$, represents a vacuum background energy.
The presence of this term has no physical consequences, and we follow
the usual practice of redefining the zero of the energy scale by
simply subtracting it out. Therefore, in terms of the effective mass
given by (51), the renormalized energy density of the ground-state of
our system is given by

\begin{equation}
\frac{\langle H_{\mbox{\tiny CGNM}}^{\mbox{\tiny M.F.}}\rangle^
{\mbox{\scriptsize vacuum}}}{L}=-\left(\frac{m^{2}}{4\pi}\right)
\left(\frac{m_{eff}}{m}\right)^{2}\left[1+2\ln 2-
\ln\left(\frac{m_{eff}}{m}\right)^{2}\right]\;.
\end{equation}

\vskip 0.5cm

\noindent Fig.1 shows the renormalized ground state energy density as
a function of effective mass $m_{eff}$. This figure reproduces the
well-know effective potential obtained in the case of $1/N$ expansion
\cite{10}.

\section{Kinetics of One-Fermion Densities}

In this section, we discuss the solutions of the nonlinear equations
of motion of the Nambu parameters $\varphi_{\bf k}$ and $\gamma_{\bf
k}$ in terms of the values taken by the parameter $d$. In order to do
this, it is useful to note that these equations can be obtained as the
canonical equations of the effective c-number Hamiltonian

\begin{equation}
h_{\mbox{\tiny{eff}}}=\frac{2}{k_0}[{\bf k}^2 + m^2(1+d)]\sigma_{\bf
k}- \frac{2m|{\bf k}|}{k_0}d \cos \gamma_{\bf k} (1-\sigma_{\bf k}^2)
^{1/2}\;,
\end{equation}

\vskip 0.3cm

\noindent where we defined $\sigma_{\bf k}=- \cos 2 \varphi_{\bf k}$
as a new momentum-like variable canonically conjugate to $\gamma_{\bf
k}$. For a general value of $d$ this dynamical system has equilibrium
solutions given by $\gamma_{\bf k}|_{\rm eq}=n\pi$ with $\varphi_{\bf
k}|_{\rm eq}$ given by Eq. (47). A special situation occurs, however,
when $d=0$. In this case, in fact, $\varphi_{\bf k}$ becomes a
constant of motion ($\gamma_{\bf k}$ becomes cyclical in
$h_{\mbox{\tiny{eff}}}$) and $\dot{\gamma}_{\bf k}$ becomes
independent of $\varphi_{\bf k}$.

Considering next the time-dependent solutions in the general case
($d\neq 0$), we see that the equilibrium solutions are stable in the
sense that small initial displacements from the equilibrium states
lead to librational behavior both in $\varphi_{\bf k}$ and in
$\gamma_{\bf k}$. For sufficiently large displacements, however, the
latter variable aquires a rotational (though nonuniform) behavior
while $\varphi_{\bf k}$ (or $\sigma_{\bf k}$) still oscillates about
an equilibrium value. A set of phase-space trajectories illustrating
this behavior is shown in Figs. 2 and 3. As one sees in these figures,
when $d$ is closer to the special value zero, the stability domains
shrink in the variable $\varphi_{\bf k}$ as the equilibrium points
approach the lines $\varphi_{\bf k}=m\pi/2$. If, on the other hand,
one considers the kinetic equations for $d=0$, one sees that the
solutions correspond to uniform $\gamma_{\bf k}$ rotations with fixed
$\varphi_{\bf k}$.

In order to interpret this behavior we recall that $d$, together with
the expansion mass $m$, and the equilibrium values of $\varphi_{\bf
k}$ and $\gamma_{\bf k}$, define the equilibrium effective mass
$m_{eff}$ [see Eq. (51)]. In particular, when $d=-1$, one has
$m_{eff}=0$ and the equations of motion reduce to

\begin{eqnarray}
\dot\varphi_{\bf k}&=&m\frac {|{\bf k}|}{k_{0}}\sin\gamma_{\bf k}
\nonumber\\ \nonumber\\ \dot\gamma_{\bf k}\sin 2\varphi_{\bf k}&=&
2 \frac {({\bf k})^{2}}{k_{0}}\sin 2\varphi_{\bf k} + 
2m\frac{|{\bf k}|}{k_{0}}\cos 2\varphi_{\bf k} \cos\gamma_{\bf k} 
\nonumber
\end{eqnarray}

\vskip 0.5cm

\noindent which reproduce the equations one obtains for a free
(massless) field e.g. by making $g=0$ in the Hamiltonian density
(27). Other values of $d$ will, on the other hand, correspond to
nonvanishing equilibrium values of the effective mass. Displacements
of $\varphi_{\bf k}$ and $\gamma_{\bf k}$ from their respective
equilibrium values will then correspond to preparing the system in a
state having a ``wrong'' (${\bf k}$-dependent) effective mass (in
particular, non-vanishing in the ``symmetric'' phase, $d=-1$). The
ensuing motion represents therefore the dynamical reaction of the
system to this situation. Several phase space trajectories for
off-equilibrium solutions are shown in Figs. 2 and 3.

The special case $d=0$ coresponds to choosing the equilibrium
effective mass $m_{eff}$ itself as the expansion mass $m$. In this
case the stability islands disappear, as the phase space trajectories
degenerate to straight lines. The off-equilibrium dynamics can then be
represented by the time dependence of the single parameter
$\gamma_{\bf k}$.

\vskip 1.0cm

\section{Discussion and conclusions}

\vskip 0.7cm

We have described a treatment of the initial-value problem in a
quantum field theory of self-interacting fermions in the Gaussian
approximation . Although the procedure is quite general, we have
implemented it for the vacuum of a relativistic many-fermion system
described by Chiral Gross-Neveu model (CGNM).

We have obtained the renormalized kinetic equations which describe the
effective dynamics of Gaussian observables in the mean-field
approximation for a uniform (1+1) dimensional system. We used the
static solution of these equations in order to renormalize the theory,
leading to an effective potential similar to that obtained by Gross
and Neveu using the $1/N$ expansion. We show also that other static
results discussed in the literature such as dynamical mass generation
due to chiral symmetry breaking and a phenomenon analogous to
dimensional transmutation can be retrieved from this formulation in
the mean-field approximation. Finally, we obtained and discussed
numerical solutions for the time evolution of the Nambu parameters
(Gaussian variables) for non-equilibrium initial conditions.

As a final comment we note that, unlike the situation found in
connection with the $1/N$ expansion, the use of the Gaussian ansatz,
Eq.(19), parametrized by the canonical transformation leading to the
quasi-fermion basis, allows for the direct dynamical determination of
the stable equilibrium situations of the system [see Eqs.(41) and
(42)], with or without symmetry breaking and mass generation. In the
latter case we obtain just a free, massless theory.

The present projection technique \cite{7} can be extended to include
and to evaluate higher dynamical correlation corrections to the
mean-field approximation. In this case the occupation numbers are no
longer constant, $\dot\nu_{{\bf k},\lambda}\neq 0$, and their time
dependence affects the effective dynamics of the Gaussian variables
(see reference \cite{9}). A finite matter density calculation beyond
the mean-field approximation allows one to study collisional
observables such as transport coefficients \cite{PIZA}.  Finally, we
comment on the extension to non-uniform systems.  In this case the
spatial dependence of the fields $\bar\psi(x)$ and $\psi(x)$ are
expanded in the natural orbitals through the use of a non-homogeneous
Bogolyubov transformation (see reference [9]).

\vskip 1.0cm

\centerline{\Large\bf Acknowledgments}

\vskip 0.7cm

One of the authors (P.L.N.) was supported by Conselho Nacional de
Desenvolvimento Cient{\'{\i}}fico e Tecnol\'ogico (CNPq), Brazil; and
by Funda{\c{c}\~ao} de Amparo \`a Pesquisa do Estado de S\~ao Paulo
(FAPESP), Brazil.
 
\vskip 1.0cm

\centerline{\Large\bf Appendix A : Uniform system and the
representations}
                      
\centerline{\Large\bf for the $\gamma^{\mu}$-matrices}

\vskip 0.7cm

A (1+1) dimensional uniform system has to be invariant under
translations and under reflection.  Reflection invariance implies that
the equations of motion are direction independent, i.e., dynamical
quantities should involve $|{\bf k}|$ only.

Let us consider the solutions of the Dirac equations
 
\begin{equation}
(i\partial\!\!\!/-m_{j})\psi^{(m_{j})}(x)=0\;,\;\;\;j=1,2\;\;,
\end{equation}
\vskip 0.5cm

\noindent for two different values $m_1$ and $m_2$ of the mass:

\begin{eqnarray}
\psi^{(m_{1})}(x)&=&\sum_{\bf k}\left(
\frac{m_{1}}{Lk_{0}^{(m_{1})}}\right)^{1/2}
\left[b_{{\bf k},1}^{(m_{1})}u_{1}^{(m_{1})}({\bf k})e^{i{\bf k}.{\bf x}}+
{b_{{\bf k},2}^{(m_{1})}}^{\dag}u_{2}^{(m_{1})}({\bf k})
e^{-i{\bf k}.{\bf x}}\right]\\\nonumber\\
\bar\psi^{(m_{1})}(x)&=&\sum_{\bf k}\left(
\frac{m_{1}}{Lk_{0}^{(m_{1})}}\right)^{1/2}
\left[{b_{{\bf k},1}^{(m_{1})}}^{\dag}\bar u_{1}^{(m_{1})}({\bf k})
e^{-i{\bf k}.{\bf x}}+
b_{{\bf k},2}^{(m_{1})}\bar u_{2}^{(m_{1})}({\bf k})
e^{i{\bf k}.{\bf x}}\right]\\ \nonumber
\end{eqnarray}

\noindent and  

\begin{eqnarray}
\psi^{(m_{2})}(x)&=&\sum_{\bf k}
\left(\frac{m_{2}}{Lk_{0}^{(m_{2})}}\right)^{1/2}
\left[b_{{\bf k},1}^{(m_{2})}u_{1}^{(m_{2})}({\bf k})
e^{i{\bf k}.{\bf x}}+
{b_{{\bf k},2}^{(m_{2})}}^{\dag}u_{2}^{(m_{2})}({\bf k})
e^{-i{\bf k}.{\bf x}}\right]\\\nonumber\\
\bar\psi^{(m_{2})}(x)&=&\sum_{\bf k}
\left(\frac{m_{2}}{Lk_{0}^{(m_{2})}}\right)^{1/2}
\left[{b_{{\bf k},1}^{(m_{2})}}^{\dag}\bar u_{1}^{(m_{2})}({\bf k})
e^{-i{\bf k}.{\bf x}}+
b_{{\bf k},2}^{(m_{2})}\bar u_{2}^{(m_{2})}({\bf k})
e^{i{\bf k}.{\bf x}}\right].\\ \nonumber
\end{eqnarray}

\noindent By imposing (say at $t=0$) the condition 

\begin{equation}
\psi^{(m_{1})}(x)=\psi^{(m_{2})}(x)
\end{equation}
\vskip  0.5cm

\noindent one can show that the operator sets $(b_{{\bf
k},1}^{(m_{1})},b_{{\bf k},2}^{(m_{1})})$ and $(b_{{\bf
k},1}^{(m_{2})},b_{{\bf k},2}^{(m_{2})})$ are related by the canonical
transformation

\begin{eqnarray}
b_{{\bf
k},1}^{(m_{2})}&=&\left\{\left(\frac{m_{1}}{m_{2}}\right)^{1/2}
\left(\frac{k_{0}^{(m_{2})}}{k_{0}^{(m_{1})}}\right)^{1/2}\bar
u_{1}^{(m_{2})} ({\bf k})u_{1}^{(m_{1})}({\bf k})\right\} b_{{\bf
k},1}^{(m_{1})}+\nonumber\\ \\
&+&\left\{\left(\frac{m_{1}}{m_{2}}\right)^{1/2}
\left(\frac{k_{0}^{(m_{2})}}{k_{0}^{(m_{1})}}\right)^{1/2} \bar
u_{1}^{(m_{2})}({\bf k})u_{2}^{(m_{1})}(-{\bf k})\right\} {b_{{-{\bf
k}},2}^{(m_{1})}}^{\dag}\nonumber\\ \nonumber\\ \nonumber\\ b_{{\bf
k},2}^{(m_{2})}&=&-\left\{\left(\frac{m_{1}}{m_{2}}\right)^{1/2}
\left(\frac{k_{0}^{(m_{2})}}{k_{0}^{(m_{1})}}\right)^{1/2}\bar
u_{2}^{(m_{1})} ({\bf k})u_{2}^{(m_{2})}({\bf k})\right\}b_{{\bf
k},2}^{(m_{1})}+ \nonumber\\ \\
&-&\left\{\left(\frac{m_{1}}{m_{2}}\right)^{1/2}
\left(\frac{k_{0}^{(m_{2})}}{k_{0}^{(m_{1})}}\right)^{1/2} \bar
u_{1}^{(m_{1})}(-{\bf k})u_{2}^{(m_{2})}({\bf k})\right\} {b_{{-{\bf
k}},1}^{(m_{1})}}^{\dag}\;\;.\nonumber\\ \nonumber
\end{eqnarray}

Reflection invariance requires that the coefficients of this canonical
transformation be invariant under ${\bf k}\rightarrow -{\bf k}$, i.e.

\begin{eqnarray}
\bar u_{1}^{(m_{2})}({\bf k})u_{1}^{(m_{1})}({\bf k})&=&\bar
u_{1}^{(m_{2})} (-{\bf k})u_{1}^{(m_{1})}(-{\bf
k})\nonumber\\\nonumber\\ \bar u_{1}^{(m_{2})}({\bf
k})u_{2}^{(m_{1})}(-{\bf k})&=&\bar u_{1}^{(m_{2})} (-{\bf
k})u_{2}^{(m_{1})}({\bf k})\nonumber\\ \\ \bar u_{2}^{(m_{1})}({\bf
k})u_{2}^{(m_{2})}({\bf k})&=&\bar u_{2}^{(m_{1})} (-{\bf
k})u_{2}^{(m_{2})}(-{\bf k})\nonumber\\ \nonumber\\ \bar
u_{1}^{(m_{1})}(-{\bf k})u_{2}^{(m_{2})}({\bf k})&=&\bar
u_{1}^{(m_{1})} ({\bf k})u_{2}^{(m_{2})}(-{\bf k})\;\;.\nonumber\\
\nonumber
\end{eqnarray}

There are just two representations of the $\gamma^{\mu}$-matrices
satisfying these conditions. These are the Pauli-Dirac representation

\begin{equation}
\gamma_{0}=\sigma_{3}\;\;\;;\;\;\;\gamma_{1}=i\sigma_{2}\;\;\;{\rm
and}\;\;\; \gamma_{5}=\gamma_{0}\gamma_{1}=\sigma_{1}
\end{equation}

\noindent and

\begin{equation}
\gamma_{0}=\sigma_{3}\;\;\;;\;\;\;\gamma_{1}=i\sigma_{1}\;\;\;{\rm
and}\;\;\; \gamma_{5}=\gamma_{0}\gamma_{1}=-\sigma_{2}\;.
\end{equation}

\noindent We choose the Pauli-Dirac representation (65) which admits
as solutions the spinors given in (29).  It should be noted, however,
that the spinors

\begin{equation}
u_{1}({\bf k})=\left(\frac{k_{0}+m}{2m}\right)^{1/2}\left[
\begin{array}{c}
         1          \\
                    \\
{\bf k}             \\
\overline{(k_{0}+m)}\\
\end{array}
\right]\;\;\;\;u_{2}({\bf k})=\left(\frac{k_{0}+m}{2m}\right)^{1/2}\left[
\begin{array}{c}
{\bf k}               \\
\overline{(k_{0}+m)}  \\
                      \\
            1
\end{array}
\right]\;\;,
\end{equation}

\vskip 0.5cm

\noindent 
which are also solutions of Dirac's equation in the Pauli-Dirac
representation, do not satisfy the uniformity conditions (64).

\vskip 1.0cm

\centerline{\Large\bf Appendix B : The uniform CGNM Hamiltonian}

In this appendix we give explicitly the Hamiltonian of the Chiral
Gross-Neveu model in a form which preserves the reflection invariance
of the system. In the Pauli-Dirac representation (see Appendix A) the
CGNM Hamiltonian is given in full by

{\footnotesize
\begin{eqnarray}
&& \hskip 4.0cm {H}_{\mbox{\tiny CGNM}}=\nonumber\\
&&\sum_{i=1}^{N}\sum_{{\bf k}'}\frac {1}{[({\bf
k}'_{i})^{2}+m^{2}]^{1/2}} \left\{({\bf k}_{i}')^{2}\left(b_{{{\bf
k}'_{i}},1}^{\dag} b_{{{\bf k}'_{i}},1}-b_{{{\bf k}'_{i}},2}b_{{{\bf
k}'_{i}},2}^{\dag}\right)- m|{\bf k}'_{i}| \left(b_{{{\bf
k}'_{i}},2}b_{{-{\bf k}'_{i}},1}+b_{{{\bf k}'_{i}},1}^{\dag} b_{{-{\bf
k}'_{i}},2}^{\dag}\right)\right\} + \nonumber\\ &&+ \left(\frac
{g^{2}m^{2}}{2L}\right)\sum_{i,j=1}^{N} \sum_{({\bf k}',{\bf k}'',{\bf
k}''',{\bf k}'''')} \left\{\frac {1}{[({\bf k}'_{i})^{2}+
m^{2}]^{1/4}[({\bf k}''_{i})^{2}+m^{2}]^{1/4}[({\bf k}'''_{j})^{2}+
m^{2}]^{1/4}[({\bf k}''''_{j})^{2}+m^{2}]^{1/4}}\right\}\times
\nonumber\\ && \left\{ b_{{{\bf k}'_{i}},1}^{\dag}b_{{{\bf
k}''_{i}},1} b_{{{\bf k}'''_{j}},1}^{\dag} b_{{{\bf
k}''''_{j}},1}\left[\xi \bar u_{1}({\bf k}'_{i}) \gamma_{5}u_{1}({\bf
k}''_{i}) \bar u_{1}({\bf k}'''_{j})\gamma_{5}u_{1}({\bf k}''''_{j})-
\bar u_{1}({\bf k}'_{i}) u_{1}({\bf k}''_{i})\bar u_{1}({\bf
k}'''_{j}) u_{1}({\bf k}''''_{j})\right] \delta_{{\bf k}'_{i}+{\bf
k}'''_{j},{\bf k}''_{i}+{\bf k}''''_{j}} \right.  \nonumber\\ &&+
\left.b_{{{\bf k}'_{i}},1}^{\dag}b_{{{\bf k}''_{i}},1} b_{{{\bf
k}'''_{j}},2} b_{{{\bf k}''''_{j}},2}^{\dag}\left[\xi\bar u_{1}({\bf
k}'_{i}) \gamma_{5}u_{1}({\bf k}''_{i}) \bar u_{2}({\bf
k}'''_{j})\gamma_{5}u_{2}({\bf k}''''_{j})- \bar u_{1}({\bf k}'_{i})
u_{1}({\bf k}''_{i})\bar u_{2}({\bf k}'''_{j})u_{2}({\bf
k}''''_{j})\right] \delta_{{\bf k}'_{i}+{\bf k}''''_{j},{\bf k}''_{i}+
{\bf k}'''_{j}}\right.\nonumber\\ &&+ \left.b_{{{\bf
k}'_{i}},1}^{\dag}b_{{{\bf k}''_{i}},1} b_{{{\bf k}'''_{j}},1}^{\dag}
b_{{{\bf k}''''_{j}},2}^{\dag}\left[\xi\bar u_{1}({\bf k}'_{i})
\gamma_{5}u_{1}({\bf k}''_{i}) \bar u_{1}({\bf
k}'''_{j})\gamma_{5}u_{2}({\bf k}''''_{j})-\bar u_{1}({\bf k}'_{i})
u_{1}({\bf k}''_{i})\bar u_{1}({\bf k}'''_{j}) u_{2}({\bf
k}''''_{j})\right] \delta_{{\bf k}'_{i}+{\bf k}'''_{j}+ {\bf
k}''''_{j},{\bf k}''_{i}}\right.\nonumber\\ &&+ \left.b_{{{\bf
k}'_{i}},1}^{\dag}b_{{{\bf k}''_{i}},1} b_{{{\bf k}'''_{j}},2}
b_{{{\bf k}''''_{j}},1}\left[\xi\bar u_{1}({\bf k}'_{i})
\gamma_{5}u_{1}({\bf k}''_{i}) \bar u_{2}({\bf
k}'''_{j})\gamma_{5}u_{1}({\bf k}''''_{j})- \bar u_{1}({\bf
k}'_{i})u_{1}({\bf k}''_{i}) \bar u_{2}({\bf k}'''_{j})u_{1}({\bf
k}''''_{j})\right] \delta_{{\bf k}'_{i},{\bf k}''_{i}+{\bf k}'''_{j}+
{\bf k}''''_{j}}\right.\nonumber\\ &&+ \left.b_{{{\bf
k}'_{i}},2}b_{{{\bf k}''_{i}},2}^{\dag} b_{{{\bf k}'''_{j}},1}^{\dag}
b_{{{\bf k}''''_{j}},1}\left[\xi\bar u_{2}({\bf k}'_{i})
\gamma_{5}u_{2}({\bf k}''_{i})\bar u_{1}({\bf
k}'''_{j})\gamma_{5}u_{1}({\bf k}''''_{j})- \bar u_{2}({\bf k}'_{i})
u_{2}({\bf k}''_{i})\bar u_{1}({\bf k}'''_{j}) u_{1}({\bf
k}''''_{j})\right] \delta_{{\bf k}'_{i}+{\bf k}''''_{j},{\bf
k}''_{i}+{\bf k}'''_{j}} \right.\nonumber\\ &&+ \left.b_{{{\bf
k}'_{i}},2}b_{{{\bf k}''_{i}},2}^{\dag} b_{{{\bf k}'''_{j}},2}
b_{{{\bf k}''''_{j}},2}^{\dag} \left[\xi\bar u_{2}({\bf
k}'_{i})\gamma_{5}u_{2}({\bf k}''_{i}) \bar u_{2}({\bf k}'''_{j})
\gamma_{5}u_{2}({\bf k}''''_{j})-\bar u_{2}({\bf k}'_{i}) u_{2}({\bf
k}''_{i}) \bar u_{2}({\bf k}'''_{j})u_{2}({\bf k}''''_{j})\right]
\delta_{{\bf k}'_{i}+{\bf k}'''_{j},{\bf k}''_{i}+{\bf
k}''''_{j}}\right.  \nonumber\\ &&+ \left.b_{{{\bf k}'_{i}},2}b_{{{\bf
k}''_{i}},2}^{\dag} b_{{{\bf k}'''_{j}},1}^{\dag} b_{{{\bf
k}''''_{j}},2}^{\dag} \left[\xi\bar u_{2}({\bf
k}'_{i})\gamma_{5}u_{2}({\bf k}''_{i}) \bar u_{1}({\bf k}'''_{j})
\gamma_{5}u_{2}({\bf k}''''_{j})-\bar u_{2}({\bf k}'_{i}) u_{2}({\bf
k}''_{i}) \bar u_{1}({\bf k}'''_{j})u_{2}({\bf k}''''_{j})\right]
\delta_{{\bf k}'_{i},{\bf k}''_{i}+{\bf k}'''_{j}+ {\bf
k}''''_{j}}\right.\nonumber\\ &&+ \left.b_{{{\bf k}'_{i}},2}b_{{{\bf
k}''_{i}},2}^{\dag} b_{{{\bf k}'''_{j}},2} b_{{{\bf k}''''_{j}},1}
\left[\xi\bar u_{2}({\bf k}'_{i})\gamma_{5}u_{2}({\bf k}''_{i}) \bar
u_{2}({\bf k}'''_{j}) \gamma_{5}u_{1}({\bf k}''''_{j})-\bar u_{2}({\bf
k}'_{i}) u_{2}({\bf k}''_{i}) \bar u_{2}({\bf k}'''_{j})u_{1}({\bf
k}''''_{j})\right] \delta_{{\bf k}'_{i}+{\bf k}'''_{j}+{\bf
k}''''_{j},{\bf k}''_{i}}\right.  \nonumber\\ &&+ \left.b_{{{\bf
k}'_{i}},1}^{\dag}b_{{{\bf k}''_{i}},2}^{\dag} b_{{{\bf
k}'''_{j}},1}^{\dag} b_{{{\bf k}''''_{j}},1} \left[\xi\bar u_{1}({\bf
k}'_{i})\gamma_{5}u_{2}({\bf k}''_{i}) \bar u_{1}({\bf k}'''_{j})
\gamma_{5}u_{1}({\bf k}''''_{j})-\bar u_{1}({\bf k}'_{i}) u_{2}({\bf
k}''_{i}) \bar u_{1}({\bf k}'''_{j})u_{1}({\bf k}''''_{j})\right]
\delta_{{\bf k}'_{i}+{\bf k}''_{i},{\bf k}'''_{j}+ {\bf
k}''''_{j}}\right.\nonumber\\ &&+ \left.b_{{{\bf
k}'_{i}},1}^{\dag}b_{{{\bf k}''_{i}},2}^{\dag} b_{{{\bf k}'''_{j}},2}
b_{{{\bf k}''''_{j}},2}^{\dag} \left[\xi\bar u_{1}({\bf
k}'_{i})\gamma_{5}u_{2}({\bf k}''_{i}) \bar u_{2}({\bf k}'''_{j})
\gamma_{5}u_{2}({\bf k}''''_{j})-\bar u_{1}({\bf k}'_{i}) u_{2}({\bf
k}''_{i}) \bar u_{2}({\bf k}'''_{j})u_{2}({\bf k}''''_{j})\right]
\delta_{{\bf k}'_{i}+{\bf k}''_{i}+{\bf k}''''_{j}, {\bf
k}'''_{j}}\right.  \nonumber\\ &&+ \left.b_{{{\bf
k}'_{i}},1}^{\dag}b_{{{\bf k}''_{i}},2}^{\dag} b_{{{\bf
k}'''_{j}},1}^{\dag} b_{{{\bf k}''''_{j}},2}^{\dag}\left[\xi\bar
u_{1}({\bf k}'_{i}) \gamma_{5}u_{2}({\bf k}''_{i}) \bar u_{1}({\bf
k}'''_{j})\gamma_{5}u_{2}({\bf k}''''_{j})- \bar u_{1}({\bf k}'_{i})
u_{2}({\bf k}''_{i})\bar u_{1}({\bf k}'''_{j}) u_{2}({\bf
k}''''_{j})\right] \delta_{{\bf k}'_{i}+{\bf k}''_{i}+{\bf k}'''_{j}+
{\bf k}''''_{j},0}\right.\nonumber\\ &&+ \left.b_{{{\bf
k}'_{i}},1}^{\dag}b_{{{\bf k}''_{i}},2}^{\dag} b_{{{\bf k}'''_{j}},2}
b_{{{\bf k}''''_{j}},1} \left[\xi\bar u_{1}({\bf
k}'_{i})\gamma_{5}u_{2}({\bf k}''_{i}) \bar u_{2}({\bf k}'''_{j})
\gamma_{5}u_{1}({\bf k}''''_{j})-\bar u_{1}({\bf k}'_{i}) u_{2}({\bf
k}''_{i}) \bar u_{2}({\bf k}'''_{j})u_{1}({\bf k}''''_{j})\right]
\delta_{{\bf k}'_{i}+{\bf k}''_{i},{\bf k}'''_{j}+ {\bf
k}''''_{j}}\right.\nonumber\\ &&+ \left.b_{{{\bf k}'_{i}},2}b_{{{\bf
k}''_{i}},1} b_{{{\bf k}'''_{j}},1}^{\dag}b_{{{\bf k}''''_{j}},1}
\left[\xi\bar u_{2}({\bf k}'_{i})\gamma_{5}u_{1}({\bf k}''_{i}) \bar
u_{1}({\bf k}'''_{j}) \gamma_{5}u_{1}({\bf k}''''_{j})-\bar u_{2}({\bf
k}'_{i}) u_{1}({\bf k}''_{i}) \bar u_{1}({\bf k}'''_{j})u_{1}({\bf
k}''''_{j})\right] \delta_{{\bf k}'_{i}+{\bf k}''_{i}+{\bf
k}''''_{j},{\bf k}'''_{j}}\right.  \nonumber\\ &&+ \left.b_{{{\bf
k}'_{i}},2}b_{{{\bf k}''_{i}},1} b_{{{\bf k}'''_{j}},2}b_{{{\bf
k}''''_{j}},2}^{\dag} \left[\xi\bar u_{2}({\bf
k}'_{i})\gamma_{5}u_{1}({\bf k}''_{i}) \bar u_{2}({\bf k}'''_{j})
\gamma_{5}u_{2}({\bf k}''''_{j})-\bar u_{2}({\bf k}'_{i}) u_{1}({\bf
k}''_{i}) \bar u_{2}({\bf k}'''_{j})u_{2}({\bf k}''''_{j})\right]
\delta_{{\bf k}'_{i}+{\bf k}''_{i}+{\bf k}'''_{j},{\bf
k}''''_{j}}\right.  \nonumber\\ &&+ \left.b_{{{\bf k}'_{i}},2}b_{{{\bf
k}''_{i}},1} b_{{{\bf k}'''_{j}},1}^{\dag} b_{{{\bf
k}''''_{j}},2}^{\dag} \left[\xi\bar u_{2}({\bf
k}'_{i})\gamma_{5}u_{1}({\bf k}''_{i}) \bar u_{1}({\bf k}'''_{j})
\gamma_{5}u_{2}({\bf k}''''_{j})-\bar u_{2}({\bf k}'_{i}) u_{1}({\bf
k}''_{i}) \bar u_{1}({\bf k}'''_{j})u_{2}({\bf k}''''_{j})\right]
\delta_{{\bf k}'_{i}+{\bf k}''_{i},{\bf k}'''_{j}+{\bf
k}''''_{j}}\right.  \nonumber\\ &&+ \left.b_{{{\bf k}'_{i}},2}b_{{{\bf
k}''_{i}},1}b_{{{\bf k}'''_{j}},2} b_{{{\bf k}''''_{j}},1}
\left[\xi\bar u_{2}({\bf k}'_{i})\gamma_{5}u_{1}({\bf k}''_{i}) \bar
u_{2}({\bf k}'''_{j}) \gamma_{5}u_{1}({\bf k}''''_{j})-\bar u_{2}({\bf
k}'_{i})u_{1}({\bf k}''_{i}) \bar u_{2}({\bf k}'''_{j})u_{1}({\bf
k}''''_{j})\right] \delta_{{\bf k}'_{i}+{\bf k}''_{i}+{\bf
k}'''_{j}+{\bf k}''''_{j},0}\right\}.  \nonumber \\\nonumber\\
\end{eqnarray}
}

This Hamiltonian is written in a particle basis.  In the dynamical
equations (35) we have to use the CGNM Hamiltonian in the Nambu
quasi-particle basis. The transformed Hamiltonian is obtained in a
straightforward way using (11) and (30).

\vskip 1.0cm

\centerline{\Large\bf Appendix C : Self-consistent renormalization}

\vskip 0.7cm

Using Eq. (46) we obtain $\sin 2\varphi_{\bf k}|_{\rm eq}$ and $\cos
2\varphi_{\bf k}|_{\rm eq}$ as

\begin{eqnarray}
\sin 2\varphi_{\bf k}|_{\rm eq}&=&\frac {-(-1)^{n}m|{\bf k}|
[1-(b+d)]}{k_{0}[k^{2}+ m^{2}(b+d)^{2}]^{1/2}}\nonumber\\ \\ \cos
2\varphi_{\bf k}|_{\rm eq}&=&\frac {[k^{2}+m^{2}(b+d)]}{k_{0}[k^{2}+
m^{2}(b+d)^{2}]^{1/2}}.\nonumber
\end{eqnarray}
\vskip 0.4cm

Substituting (69) in $I_{1}$ and $I_{2}$ given in (37), taking
$\nu_{{\bf k},1}=\nu_{{\bf k},2}=0$, and performing the integration,
we have

{\footnotesize
\begin{eqnarray}
I_{1}\!&=&\!\int_{-\Lambda}^{+\Lambda} \frac {dk}{(k^{2}+m^{2})}\frac
{[{\bf k}^{2}+ m^{2}(b+d)]}{[k^{2}+m^{2}(b+d)^{2}]}=a+ b\ln\left(\frac
{\Lambda^{2}}{m^{2}}\right) \nonumber\\ \nonumber\\ \nonumber\\
a\!&=&\!\frac{2d}{[(1+d)^{2}-1]^{1/2}}\arctan [(1+d)^{2}-1]^{1/2}+
\ln\left[\frac {4}{(1+d)^{2}}\right]\;\;{\rm and}\;\;b\;=\;1
\;\;\nonumber\\ &&{\rm when}\;\;(1+d)^{2}>1\nonumber\\ \\
a\!&=&\!\frac{d}{[1-(1+d)^{2}]^{1/2}} \ln\left[\frac
{1+\{1-(1+d)^{2}\}^{1/2}} {1-\{1-(1+d)^{2}\}^{1/2}}\right]+
\ln\left[\frac {4}{(1+d)^{2}}\right]\;\;{\rm and}\;\;b\;=\;1
\;\;\nonumber\\ &&{\rm when}\;\;(1+d)^{2}<1\nonumber
\end{eqnarray}
}
 
{\footnotesize
\begin{eqnarray}
I_{2}\!&=&\!(b+d-1)\int_{-\Lambda}^{\Lambda}\frac {dk}{(k^{2}+m^{2})}
\frac{k^{2}}{[k^{2}+m^{2}(b+d)^{2}]^{1/2}}=
c+d\ln\left(\frac{\Lambda^{2}}{m^{2}}\right) \nonumber\\ \nonumber\\
\nonumber\\
c\!&=&\!\frac{-2d}{[(1+d)^{2}-1]^{1/2}}\arctan[(1+d)^{2}-1]^{1/2}+
d\ln\left[\frac{4}{(1+d)^{2}}\right]\;\;{\rm and}\;\;d\;=\;d
\;\;\nonumber\\ &&{\rm when}\;\;(1+d)^{2}>1\nonumber\\ \\
c\!&=&\!\frac {-d}{[1-(1+d)^{2}]^{1/2}} \ln\left[\frac
{1+\{1-(1+d)^{2}\}^{1/2}}{1-\{1-(1+d)^{2}\}^{1/2}}\right]+
d\ln\left[\frac{4}{(1+d)^{2}}\right]\;\;{\rm and}\;\;d\;=\;d
\;\;\nonumber\\ &&{\rm when}\;\;(1+d)^{2}<1\nonumber
\end{eqnarray}
}

This shows that the ansatz (44)-(45) is self-consistent. Eqs. (70) and
(71) give the values of the constants $a$, $b$ and $c$. We observe
that $d$ remains arbitrary in this calculation.

\newpage

\centerline{\Large\bf Figure Captions}

\vskip 0.7cm

Figure 1 - Mean-field effective potential for the ground-state of an
uniform fermion system described by the CGNM as function of effective
mass $m_{eff}$.

\vskip 0.7cm

Figure 2 - Phase-space of Nambu parameters for 
$|{\bf k}|=m=\sqrt{2}$ in the case of free 
fermion system [$(d+1=0)$ or $m_{eff}=0$]. 
Initial-values: \\
Dotted line : $\varphi_{{\bf k}}^{\rm in}=
(-0.2+n\pi/2)\;,\;\gamma_{\bf k}^{\rm in}=2n\pi$ and 
$\varphi_{{\bf k}}^{\rm in}=(+0.2+n\pi/2)\;,\;
\gamma_{\bf k}^{\rm in}=(2n+1)\pi$, \\
Dashed line : $\varphi_{{\bf k}}^{\rm in}=
(-0.1+n\pi/2)\;,\;\gamma_{\bf k}^{\rm in}=2n\pi$ and 
$\varphi_{{\bf k}}^{\rm in}=(+0.1+n\pi/2)\;,\;
\gamma_{\bf k}^{\rm in}=(2n+1)\pi$, \\
Solid line : $\varphi_{{\bf k}}^{\rm in}=
(-\pi/4+n\pi/2)\;,\;\gamma_{\bf k}^{\rm in}=0$ and 
$\varphi_{{\bf k}}^{\rm in}=(0+n\pi/2)\;,\;
\gamma_{\bf k}^{\rm in}=0$, \\
Dot-Dashed line : $\varphi_{{\bf k}}^{\rm in}=
(\pi/8+n\pi/2)\;,\;\gamma_{\bf k}^{\rm in}=0$.

\vskip 0.7cm

Figure 3 - Same as Fig. 2 with $(d+1)=0.5858$. Initial values: \\
Dotted line : $\varphi_{{\bf k}}^{\rm in}=
(-0.1+n\pi/2)\;,\;\gamma_{\bf k}^{\rm in}=2n\pi$ and 
$\varphi_{{\bf k}}^{\rm in}=(+0.1+n\pi/2)\;,\;
\gamma_{\bf k}^{\rm in}=(2n+1)\pi$, \\
Dashed line : $\varphi_{{\bf k}}^{\rm in}=
(-0.05+n\pi/2)\;,\;\gamma_{\bf k}^{\rm in}=2n\pi$ and 
$\varphi_{{\bf k}}^{\rm in}=(+0.05+n\pi/2)\;,\;
\gamma_{\bf k}^{\rm in}=(2n+1)\pi$, \\
Solid line : $\varphi_{{\bf k}}^{\rm in}=
(-\pi/8+n\pi/2)\;,\;\gamma_{\bf k}^{\rm in}=0$ and 
$\varphi_{{\bf k}}^{\rm in}=(0+n\pi/2)\;,\;
\gamma_{\bf k}^{\rm in}=0$, \\
Dot-Dashed line : $\varphi_{{\bf k}}^{\rm in}=
(3\pi/16+n\pi/2)\;,\;\gamma_{\bf k}^{\rm in}=0$. 


\vskip 1.0cm

\begin{thebibliography}{9}

\bibitem{1} M. Samiullah, O. \'Eboli and 
S.-Y. Pi, Phys. Rev. {\bf D 44}, 2335 (1991), 
and references therein;

            D. Boyanovski, H. J. de Vega 
and R. Holman, Phys. Rev. {\bf D 49}, 2769 (1994);

            F. L. Braghin, C. Martin and D. Vautherin, 
Phys. Lett. {\bf B 348}, 343 (1995).

\bibitem{2} M. Ploszajczak and M. Rhodes-Brown,  
Phys. Rev. {\bf D 33}, 3686 (1986); 

            H.-Th. Else, M. Gyulassy and D. Vasak, 
Phys. Lett. {\bf B 177}, 402 (1986);

            C. Miug Ko, Qi Li and Renchuan Wang, 
Phys. Rev. Lett. {\bf 59}, 1084 (1987).

\bibitem{3} M. C. Nemes and A. F. R. de Toledo Piza, 
Physica {\bf A 137}, 367 (1986), 
and references therein. 

\bibitem{4} O. \'Eboli, S.-Y. Pi and R. Jackiw, 
Phys. Rev. {\bf D 37}, 3557 (1988);

            A. Kovner and B. Rosenstein, 
Ann. Phys. (N.Y.) {\bf 187}, 449 (1988);

            R. Jackiw, Physica {\bf A 158}, 269 (1989).  

\bibitem{KeLi} A. K. Kerman and D. Vautherin, Ann. Phys. (N.Y.) {\bf
192}, 408 (1989); 

               A. K. Kerman and C. Y. Lin, Ann. Phys. (N.Y.) {\bf
241}, 185 (1995). 

\bibitem{5} J. des Cloiseaux, in {\it The Many-Body Physics}, 
edited by C. de Witt and 
R. Balian (Gordon and Breach, N. Y., 1968).

\bibitem{6} Shan-Jin Chang, Phys. Rev. {\bf D 12}, 1071 (1975).

\bibitem{7} A. F. R. de Toledo Piza, 
in {\it Time-Dependent Hartree-Fock and Beyond}, 
edited by K. Goeke and P.-G. Reinhardt, Lectures Notes in Physics 171 
(Springer-Verlag, Berlin, 1982); 

            M. C. Nemes and A. F. R. de Toledo Piza, 
Phys. Rev. {\bf C 27}, 862 (1983);

            B. V. Carlson, M. C. Nemes and A. F. R. de Toledo Piza, 
Nucl. Phys {\bf A 457}, 261 (1986).

\bibitem{8} P. Buck, H. Feldmeier and M. C. Nemes, 
Ann. Phys. (N.Y.) {\bf 185}, 170 (1988).

\bibitem{9} L. C. Yong and A. F. R. de Toledo Piza, 
Phys. Rev. {\bf D 46}, 742 (1992); 

            L. C. Yong, Doctoral Thesis, University of S\~ao Paulo, 
1991 (unpublished). 

\bibitem{10} D. J. Gross and A. Neveu, Phys. Rev. {\bf D 10}, 3235 (1974). 

\bibitem{11} P. L. Natti, Doctoral Thesis, University of S\~ao Paulo, 
1995 (unpublished).

\bibitem{12} A. Kerman and T. Troudet, 
Ann. Phys. (N.Y.) {\bf 154}, 456 (1984).

\bibitem{DHN} R. Daschen, B. Hasslacher and A. Neveu, 
Phys. Rev. {\bf D 12}, 2443 (1975).

\bibitem{NJ} Y. Nambu and G. Jona-Lasinio, 
Phys. Rev. {\bf 122}, 345 (1961).

\bibitem{EW} E. Witten, Nucl. Phys. {\bf B 145}, 110 (1978). 

\bibitem{BKKW} B. Berg, M. Karowski, V. Kurac and P. Weisz, Nucl. Phys. 
{\bf B 134}, 125 (1978). 

\bibitem{TOM} P. R. I. Tommasini, Doctoral Thesis, University of 
S\~ao Paulo, 1995 (unpublished). 
                           
\bibitem{RK} B. Rosenstein, and A. Kovner,  Phys. Rev. {\bf D40}, 523 (1989);

             R. Pausch, M. Thies and V. L. Dolman, 
Z. Phys. {\bf A338}, 441 (1991).

\bibitem{PIZA} A.F.R. de Toledo Piza, Proc of the IX 
Workshop in Nuclear Physics, Buenos Aires, 
Argentina, World Scientific (1987).

\end{thebibliography}
\end {document}
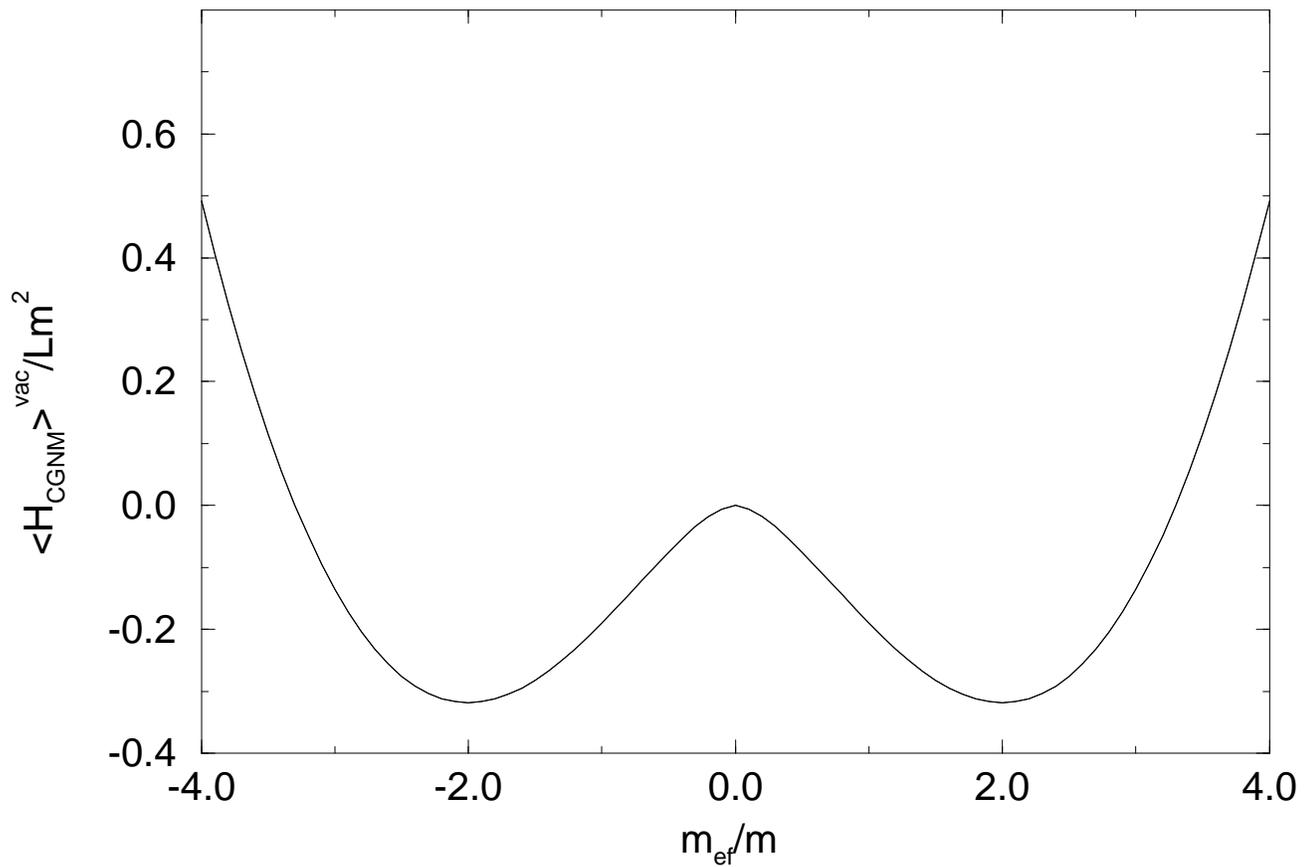
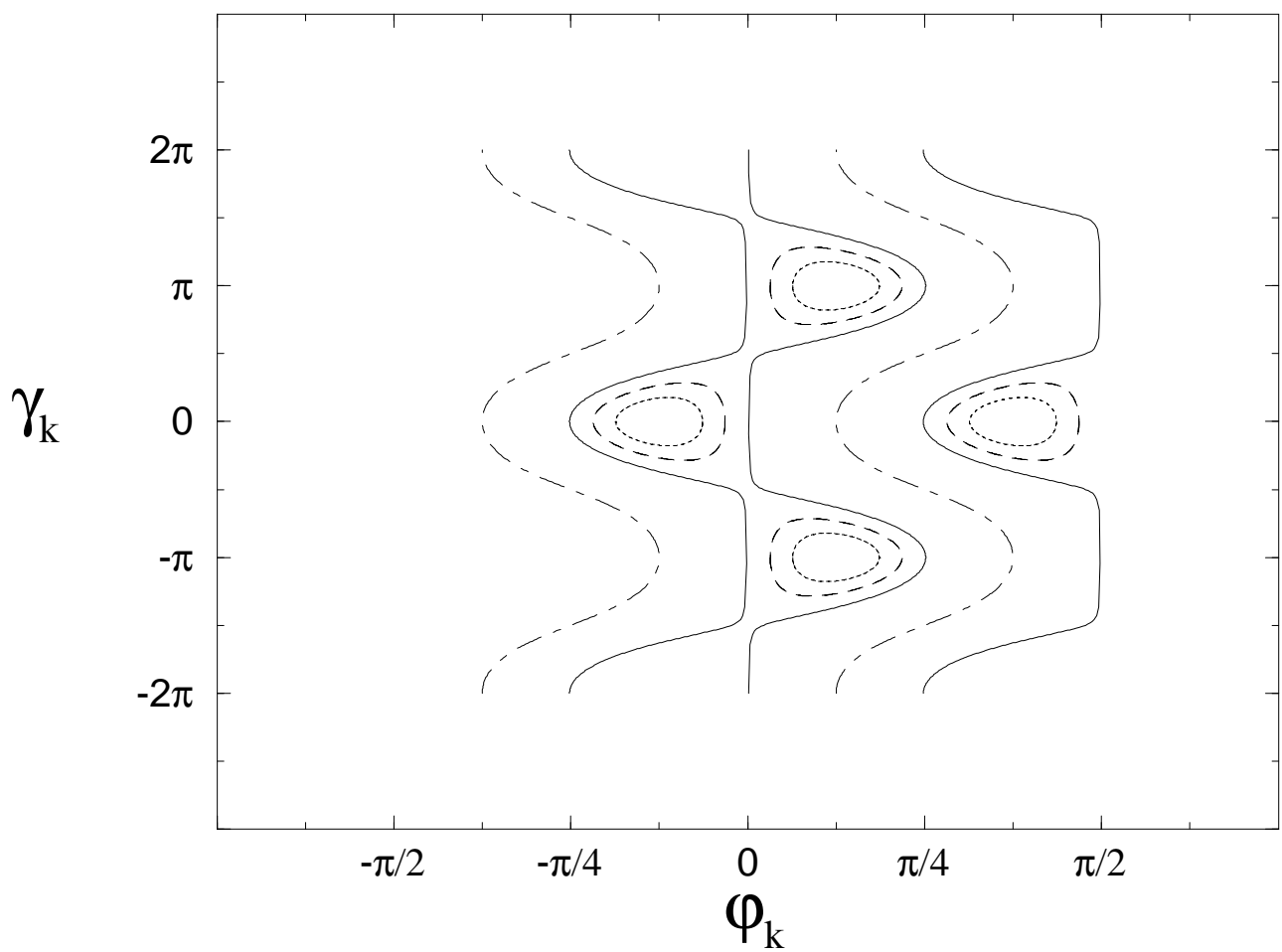
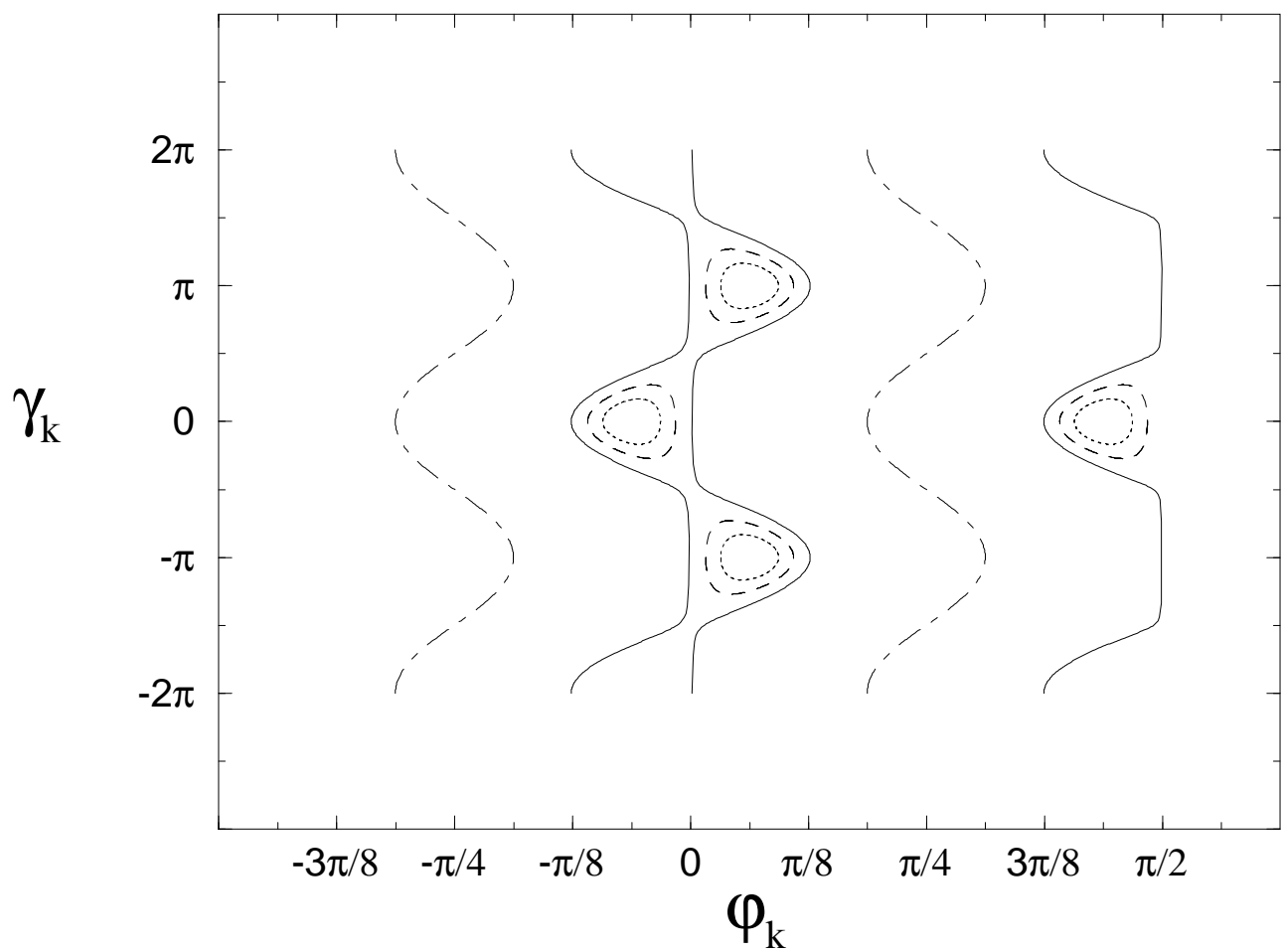
